\documentclass[10pt,journal,compsoc]{IEEEtran}
\usepackage{amsmath,amsthm,amsfonts,latexsym,amssymb,bbm}
\usepackage{graphicx,xcolor}
\usepackage{algorithmic,algorithm,subfigure}
\usepackage{ifthen}

\ifCLASSOPTIONcompsoc
  \usepackage[nocompress]{cite}
\else
  \usepackage{cite}
\fi
\usepackage{graphicx,xcolor}
\ifCLASSINFOpdf
\else
\fi
\usepackage{amsmath,amsthm,amsfonts,latexsym,amssymb,bbm}
\usepackage{subfigure}

\newcommand{\bOne}{\mathbf 1}
\newcommand{\Reals}{\mathbb R}
\newcommand{\Integers}{\mathbb Z}
\newcommand{\IntegersP}{\Integers_+}

\newcommand{\Prob}{\mathbb{P}}
\newcommand{\E}{\mathbb{E}}

\newtheorem{allcnt}{AllCnt}[section]

\newtheorem{theorem}[allcnt]{Theorem}
\newtheorem{corollary}[allcnt]{Corollary}
\newtheorem{lemma}[allcnt]{Lemma}

\newboolean{showcomments}
\setboolean{showcomments}{true}
\newcommand{\comment}[1]{  \ifthenelse{\boolean{showcomments}}
{ \textcolor{red}{#1}} {}  }

\begin{document}
%
\title{Asynchronous Approximation of a Single Component of the Solution to a Linear System}
%
%
%
%

\author{Asuman~Ozdaglar,~
        Devavrat~Shah,~
				and~Christina~Lee~Yu~
\IEEEcompsocitemizethanks{
\IEEEcompsocthanksitem Asuman Ozdaglar and Devavrat Shah are professors at Massachusetts Institute of Technology in the Electrical Engineering and Computer Science Department.
\IEEEcompsocthanksitem Christina Lee Yu is an assistant professor at Cornell University in the School of Operations Research and Information Engineering. e-mail: cleeyu@cornell.edu.
}
}

\IEEEtitleabstractindextext{%
\begin{abstract}
We present a distributed asynchronous algorithm for approximating a single component of the solution to a system of linear equations $Ax = b$, where $A$ is a positive definite real matrix and $b \in \Reals^n$. This can equivalently be formulated as solving for $x_i$ in $x = Gx + z$ for some $G$ and $z$ such that the spectral radius of $G$ is less than 1. Our algorithm relies on the Neumann series characterization of the component $x_i$, and is based on residual updates. We analyze our algorithm within the context of a cloud computation model motivated by frameworks such as Apache Spark, in which the computation is split into small update tasks performed by small processors with shared access to a distributed file system. We prove a robust asymptotic convergence result when the spectral radius $\rho(|G|) < 1$, regardless of the precise order and frequency in which the update tasks are performed. 
We provide convergence rate bounds which depend on the order of update tasks performed, analyzing both deterministic update rules via counting weighted random walks, as well as probabilistic update rules via concentration bounds. The probabilistic analysis requires analyzing the product of random matrices which are drawn from distributions that are time and path dependent. We specifically consider the setting where $n$ is large, yet $G$ is sparse, e.g., each row has at most $d$ nonzero entries. This is motivated by applications in which $G$ is derived from the edge structure of an underlying graph. Our results prove that if the local neighborhood of the graph does not grow too quickly as a function of $n$, our algorithm can provide significant reduction in computation cost as opposed to any algorithm which computes the global solution vector $x$. Our algorithm obtains an $\epsilon \|x\|_2$ additive approximation for $x_i$ in constant time with respect to the size of the matrix when the maximum row sparsity $d = O(1)$ and $1/(1-\|G\|_2) = O(1)$, where $\|G\|_2$ is the induced matrix operator 2-norm.
\end{abstract}

}

\maketitle

\IEEEdisplaynontitleabstractindextext

%
\IEEEpeerreviewmaketitle

\IEEEraisesectionheading{\section{Introduction}\label{sec:introduction}}

%
%
%
%
\IEEEPARstart{I}{magine}
that you are a small restaurant owner in a city. You would like to obtain a quantitative estimate of how your popularity and reputation compare to your competitors within a 5 mile radius of you. You may want to compare the significance of the associated websites of your restaurant and other similar restaurants within the webgraph. This can be measured by PageRank, a quantity used by Google to rank search results. PageRank is defined as the solution to $x = \alpha \bOne / n + (1 - \alpha) P^T x$, where $P$ is the adjacency matrix of the webgraph, $\alpha$ is a given parameter, $\bOne$ is the vector of all ones, and $n$ is the dimension. Alternatively, you may want to compare the social influence of the restaurants' associated Facebook pages, which can be computed via the Bonacich centrality. Bonacich centrality is defined as the solution to $x = (I - \alpha G)^{-1} \bOne$, where $G$ is the adjacency matrix of the social network, and $\alpha$ is a given parameter. Both PageRank and Bonacich centrality can be formulated as the solution to a system of linear equations, where the dimension is as large as the webpages in the webgraph or the number of Facebook pages, which is an overwhelming computational expense for our hypothetical small restaurant owner. In this paper, we investigate the question: can we obtain estimates of a few coordinates of the solution vector without the expense of approximating the entire solution vector?

We consider approximating the $i^{\text{th}}$ component of the solution to a linear system of equations $Ax = b$,
where $A$ is a positive definite $n \times n$ real matrix, and $b$ is a vector in $\Reals^n$. Positive definite matrices include symmetric diagonally dominant matrices, such as the Laplacian, and also our motivating examples of network centralities, PageRank and Bonacich centrality. Note that $A$ or $G$ may not be symmetric. When $A$ is positive definite, there exists a choice of $G$ and $z$ such that the problem is equivalent to approximating the $i^{\text{th}}$ component of the solution to $x = Gx + z$, and the spectral radius of $G$, denoted $\rho(G)$, is less than 1. For PageRank, $\rho(G)$ is a constant, bounded by the teleportation probability, independent of the underlying graph. For Bonacich centrality, $\rho(G)$ can be chosen to be less than 1 by a proper choice of the ``discount factor'' for any graph. 

We consider a setting with large $n$ and sparse $G$, i.e., the number of nonzero entries in every row of $G$ is at most $d$. This form of sparsity arises when the matrix is derived from an underlying bounded degree graph. We will also discuss how we can relax this constraint to graphs for which the local neighborhood size does not grow too quickly.

Solving large systems of linear equations is a problem of great interest due to its relevance to a variety of applications across science and engineering, such as solving large scale optimization problems, approximating solutions to partial differential equations, and modeling network centralities. Due to the large scale of these systems, it becomes useful to have an algorithm which can approximate only a few components of the solution without computing over the entire matrix. Such an algorithm would also lead to efficient ranking and comparison methods. As solving a system of linear equations is fundamentally a problem which involves the full matrix, computing a single component of the solution is non-trivial. 

In this era of big data, the classic computation model has changed significantly to accomodate for computation which is too large to compute within a single processor's memory. We will consider a distributed cloud computation model inspired by frameworks such as Mapreduce \cite{DeanGhemawat08} or its open source implementation Hadoop \cite{Borthakur07} 
or its memory efficient open-source implementation Spark \cite{ZahariaChowdhuryFranklinShenkerStoica10}, in which there are many 
processors with small constant size memory, yet they have access through the cloud  to a distributed file 
system (DFS) 
which stores the information regarding the entire matrix. Our algorithm will consist of a sequence of small tasks which can be assigned to different processors to compute asynchronously. We will measure the cost of our algorithm via the amount of computational resources consumed, e.g. number of tasks, DFS accesses, and memory consumed.

\subsection{Problem Statement and Notation}

Given an index $i \in [n]$, a vector $z \in \Reals^n$, and a matrix $G \in \Reals^{n \times n}$ such that $\rho(G) < 1$, the goal is to solve for $x_i$, the $i$-th component of the solution vector to $x = Gx + z$. Throughout the paper, we associate a graph to the matrix $G$, and we will provide our analysis as a function of properties of the graph. Let $\mathcal{G}(G) = (\mathcal{V,E})$ denote the directed graph where $\mathcal{V} = [n]$, and $(u,v) \in \mathcal{E}$ if and only if $G_{uv} \neq 0$. 
Each coordinate of vector $x$ corresponds to a vertex in $\mathcal{V}$. Let $N_u(t) \subset \mathcal{V}$ denote the vertices with path length $t$ from vertex $u$, specifically $v \in N_u(t)$ if there exists a path from $u$ to $v$ of length $t$, allowing for loops and repeated visits to vertices. We denote the immediate neighbors of vertex $u$ by $N_u$, i.e., $v \in N_u$ if $G_{uv} \neq 0$. The sparsity assumption on $G$ means that $|N_u| \leq d$ for all $u$. 
We summarize notation used in the paper:

\medskip
\begin{center}
\begin{tabular}{| r l |}
\hline
$\rho(G)$ & the spectral radius of $G$ \\
$\tilde{G}$ & the matrix s.t. $\tilde{G}_{ij} = |G_{ij}|$ \\
$\|r\|_0$ & the sparsity of vector $r$ \\
$\|r\|_p$ & the vector $p$-norm for $p \geq 1$ \\
$\|G\|_p$ & the induced matrix operator norm \\
$\mathcal{E}$ & $\{(u,v) ~s.t.~ G_{uv} \neq 0\}$ \\
$N_u$ & $\{v \in [n] ~s.t.~ G_{uv} \neq 0\}$ \\
$N_u(t)$ & vertices with path length $t$ from $u$ \\
$d$ & the maximum degree, $\max_u |N_u|$ \\
$r^{(t)}$ & residual $(G^T)^t e_i$ \\
$\hat{x}_i^{(t)}$  & estimate $z^T \sum_{k=0}^{t-1} (G^T)^k e_i$ \\
\hline
\end{tabular}
\end{center}


\subsection{Equivalence of $Ax = b$ and $x = Gx + z$} \label{sec:Axb}

Given a system of linear equations of the form $Ax = b$, there are many methods for choosing $G$ and $z$ such that the equation is equivalent to the form given by $x = Gx + z$ with $\rho(G) < 1$ \cite{Westlake68}. The Jacobi and Richardson methods are suitable for our setting because they have additional properties that $G$ is as sparse as $A$, and $G_{ij}$ can be computed as a simple function of $A_{ij}$ and $A_{ii}$. Given $(A, b)$, there may be many ways to choose $(G, z)$ to satisfy the condition that $\rho(G) < 1$. Finding the optimal choice\footnote{By optimal, we would like to minimize $\rho(G)$, which maximizes the convergence rate of the algorithm.} of $(G, z)$ given $(A,b)$ is beyond the scope of this paper.

\begin{corollary} \label{corr:Axb_xGxz}
If $A$ is positive definite or diagonally dominant, we can use standard methods (e.g. Jacobi or Richardson), to choose $(G,z)$ such that $\rho(G) < 1$, and the solution $x$ which satisfies $x = Gx + z$ will also satisfy $A x = b$.
\end{corollary}

The Jacobi method chooses $G = - D^{-1} (A - D)$ and $z = D^{-1} b$, where $D$ is a diagonal matrix such that $D_{uu} = A_{uu}$. The Richardson method chooses $G = I - \gamma A$ and $z = \gamma b$ for any $\gamma$ such that $0 \leq \gamma \leq \min_{\|x\|_2 = 1} (2 x^T A x) / (x^T A^T A x)$. If $A$ is symmetric, then using the Richardson method with an optimal choice of $\gamma$ results in a choice of $G$ such that $\rho(G) = \|G\|_2 = (\kappa(A) - 1)/(\kappa(A) + 1)$, where $\kappa(A)$ denotes the condition number of $A$. 

\subsection{Contributions and Summary of Results}


We introduce novel algorithms and corresponding analyses for estimating $\hat{x}_i$ for a single component $i$ of the solution vector to $x = Gx + z$. Our algorithm can be implemented in a fully distributed, asynchronous computation model suitable for Mapreduce / Hadoop or Spark (discussed in Section \ref{sec:model}). For large yet sparse systems ($n$ large but $d$ is bounded or grows very slowly with $n$), the computational cost of our algorithms are significantly less than global algorithm which incur the cost of computing the full solution vector $x$. In addition our algorithms improve upon Monte Carlo methods for single component analysis which exhibit high variance and slow convergence.


\medskip{\bf Algorithm.} 
Our algorithm  relies upon the Neumann series representation of the solution, i.e. $x = \sum_{k=0}^{\infty} G^k z$,
\begin{align}
x_i = e_i^T \textstyle\sum_{k=0}^{\infty} G^k z = z^T \textstyle\sum_{k=0}^{\infty} (G^T)^k e_i, \label{eq:neumann}
\end{align}
where $e_i$ denotes the standard basis vector which takes value 1 at coordinate $i$ and 0 elsewhere. We can interpret the term $z^T (G^T)^k e_i$ to be the weighted sum of all walks of length $k$ from vertex $i$ on the graph defined by $G$. Since we focus on approximating only $x_i$, we can compute the $k$ lower order terms of the summation by summing weighted walks within the $k$-radius neighborhood of vertex $i$, as opposed to the entire graph. This introduces a locality in computation that we can exploit if the neighborhoods of vertex $i$ do not grow quickly.


Our algorithm is an iterative residual based method in which every task corresponds to updating one coordinate of the residual vector. Let us define residual vectors $r^{(t)} = (G^T)^t e_i$. For any $t \in \Integers$, expression \eqref{eq:neumann} can be rearranged as
\begin{align}
x_i 
&= z^T \textstyle\sum_{k=0}^{t-1} r^{(k)} + \left(z^T \textstyle\sum_{k=0}^{\infty} (G^T)^k \right) (G^T)^t e_i \nonumber \\
&= z^T \textstyle\sum_{k=0}^{t-1} r^{(k)} + x^T r^{(t)}. 
\end{align}
At iteration $t$, the algorithm estimates according to $\hat{x}_i^{(t)} = z^T \textstyle\sum_{k=0}^{t-1} r^{(k)}$, and the estimation error will be $x^T r^{(t)}$. The synchronous implementation of the algorithm updates the estimate by adding the value of the residual vector in each iteration and updating the residual by multiplying by $G^T$. The algorithm terminates when $\|r^{(t)}\|_2 < \epsilon$, which guarantees that $|\hat{x}_i^{(t)} - x_i| \leq \epsilon \|x\|_2$.

\medskip{\bf Asynchronous Implementation.}
The asynchronous implementation of the algorithm updates one coordinate of the residual vector at a time. Updating coordinate $u$ corresponds to adding $r_u$ to $\hat{x}_i$, and multiplying $r_u$ by the $u^{\text{th}}$ row of $G$ and adding that to the residual vector $r$. These updates can be interpreted as accumulating weights of walks over the graph, beginning with short length walks. Every update task maintains an invariant
\[x_i = \hat{x}_i + r^T x,\]
where $\hat{x}_i$ denotes the estimate, and $r$ denotes the residual vector. The invariant property characterizes the error at every iteration, which is used to prove the algorithm always converges when $\rho(\tilde{G}) < 1$, where $\tilde{G} = |G|$, i.e. $\tilde{G}_{ij} = |G_{ij}|$ for all $i,j$. The convergence holds regardless of the order in which coordinates are updated, as long as each coordinate is updated infinitely often. It is robust to asynchronous updates in which the computation corresponding to different tasks may interweave in the order they update the residual vector in the DFS. The conditions are given in terms of matrix $\tilde{G}$ rather than $G$, since the asynchronous implementation may sum walks of different lengths simultaneously. We use $\tilde{G}^l$ to obtain a worst case bound on the total contribution of any set of walks of length longer than $l$. We do not require uniform bounds on communication delays or clock rates, as often needed for similar results in asynchronous computation (see \cite{BertsekasTsitsiklis89}). 

\medskip{\bf Computational Cost.}
Our algorithm requires $O(n + |\mathcal{E}|)$ space in the distributed file system, and a single update task requires $O(|N_u|)$ DFS accesses, where $u$ is the coordinate being updated. The convergence rate of our algorithm can be analyzed via the evolution of the residual vector $r$, which is a function of the particular order, or sequence, of tasks in which the coordinates are updated. The sparsity pattern of the residual vector will grow according to an expanding local neighborhood around vertex $i$ in the graph defined by $G$, allowing us to upper bound the number of update tasks needed by the computation as a function of the size of this neighborhood. 
We analyze different implementations of our algorithm, corresponding to variants for choosing the order for updating coordinates of the residual vector. We provide our bounds as a function of the maximum degree of the graph, denoted by $d$, but we can extend the results to other graphs in which we have an upper bound for how the size of the local neighborhood grows.

As a baseline, we compute the cost of a synchronous distributed implementation in which the tasks coordinate between iterations to update the residual vector according to $r^{(t)} = G r^{(t-1)}$, which involves $\|r^{(t)}\|_0$ individual coordinate update tasks. We prove that the synchronous implementation attains error less than $\epsilon \|x\|_2$ with at most
\[O(\min(\epsilon^{\ln(d)/\ln(\|G\|_2)}, n\ln(\epsilon)/\ln(\|G\|_2)))\]
update tasks. This calculation assumes that the computation is synchronized across iterations.

We analyze the asynchronous implementation in which the update tasks do not coordinate different iterations of computation, but rather update the same residual vector $r$ in the DFS. Rather than multiplying by matrix $G$ in each iteration, every individual update task corresponds to applying a local update involving a single row of the matrix $G$. When the coordinates update sequentially in the order imposed by the expanding local neighborhoods of vertex $i$, the convergence rate is very similar to the synchronous implementation, requiring at most 
\[O(\min(\epsilon^{\ln(d)/\ln(\|\tilde{G}\|_2)}, n\ln(\epsilon)/\ln(\|\tilde{G}\|_2)))\]
update tasks until the error is less than $\epsilon \|x\|_2$. This update rule ensures that first all coordinates in $N_i(1)$ are updated, followed by all coordinates in $N_i(2)$, where the coordinates within the same neighborhood can be updated in any order. The order of updates can be coordinated by a designated master processor which manages a shared task queue for all other processors. This update order ensures that short walks get counted in the estimate earlier. The bound depends on $\tilde{G}$ due to using a worst case upper bound for the weight of all walks of length longer than a certain value. Compared to the synchronous implementation, this analysis is weaker when $G$ may have positive or negative entries, since $\|G\|_2 \leq \|\tilde{G}\|_2$.

We can alternatively employ randomness to sample the next coordinate to update, enabling every processor to generate the next update task without any coordination cost among other tasks. The algorithm adaptively samples the next coordinate to update according to a distribution which depends on the current residual vector. When the sequence of coordinate updates are sampled uniformly amongst coordinates with nonzero residual values, we can guarantee that with probability at least $1 - \delta$, the error contracts by a time varying factor in each step, and the algorithm involves at most
\[O(\min((\epsilon \sqrt{\delta/5})^{-d/(1-\|G\|_2)}, -n \ln (\epsilon \sqrt{\delta})/(1-\|G\|_2)))\]
update tasks until the error is less than $\epsilon \|x\|_2$. We term this `uniform censored sampling', since we censored the coordinates according to whether the residual value is nonzero, and we sample uniformly otherwise. Establishing the convergence rate requires bounding the Lyapunov exponent for a product of random matrices drawn from time and path dependent distributions. This is inherently different from previous analyses of randomized coordinate updates, which sample from a history independent distribution. We developed a new analysis for `uniform censored sampling' updates.

We can compare with the bounds for the synchronous implementation by considering that $1 - \|G\|_2 \approx - \ln(\|G\|_2)$ when $\|G\|_2 \approx 1$. The randomized update implementation scales exponentially with $d$, whereas the other two bounds only scale polynomially with $d$. The gap is due to the fact that the synchronous and deterministic asynchronous implementations update in an order which ensures that short walks are counted earlier. Intuitively, we expect that the weight of the walks decays exponentially due to the weight being a product over values in $G$ which converge eventually to zero. Therefore, by sampling uniformly amongst all coordinates with nonzero residuals, the algorithm may choose to update coordinates which are farther away from vertex $i$ before it finishes updating coordinates within a closer neighborhood of $i$. As a result of a single update task, the contributions added in the process corresponding to updates of the residuals along neighbors will will be approximately ``exponentially less significant'', and yet the coordinates still carry an equal weight in determining the next coordinate to update. This leads to the exponentially slower convergence as a function of $d$. This can be remedied by emphasizing coordinates with larger residuals, which we explore heuristically through simulations.

The right hand expressions within the convergence rate bounds across the different implementations are essentially the same, and provide a comparison of our algorithm to standard linear iterative methods, which also converge at the same rate. The left hand expressions provide a local analysis utilizing the sparsity of $G$. They show that the number of tasks required by our algorithm to reach a specified precision is constant with respect to $n$ as long as $d = O(1)$ and $1/(1-\|G\|_2) \approx -1/\ln(\|G\|_2) = O(1)$. The analysis shows that as long as the local neighborhood does not grow too quickly, i.e., the network is large and sparse enough, and the spectral properties are well behaved, i.e., $\|G\|_2$ is bounded away from 1, there is a $n_0$ such that for all $n \geq n_0$, our algorithm obtains an estimate of $x_i$ with fewer computational tasks than any centralized algorithm, by the simple fact that the required tasks of our algorithm is upper bounded by an expression which is independent of $n$, and any centralized algorithm will scale at least as the size of the solution vector.

\subsection{Related Work}

There are not many existing methods which have explored single component approximations of the solution vector. Most standard techniques such as Gaussian elimination, factorization or decomposition, gradient methods, and linear iterative methods all compute the full solution vector, and thus the computation involves all coordinates and all entries in the matrix \cite{Westlake68, GolubVanLoan13}. Most of the methods are either stationary linear iterative methods (e.g. Jacobi, Gauss-Seidel, successive over-relaxations) or optimization algorithms for minimizing $\frac12 x^T A x - b^T x$ or $\|A x - b\|_2^2$. For example, Kaczmarz, Gauss-Seidel, or Gauss-Southwell are all variations of either coordinate descent or gradient descent \cite{Wright15}. 

Stationary linear iterative methods use updates of the form $x_{t+1} = G x_t + z$ to recursively approximate leading terms of the Neumann series. The error after $t$ iterations is given by $G^t (x - x_0)$, thus the number of iterations to achieve $\|x_t - x\|_2 \leq \epsilon \|x\|_2$ is $\ln(\epsilon)/\ln(\|G\|_2)$. For any $t$, $x_t$ will be at least as dense as $z$, and there is no reason to assume $z$ is sparse; a single update step could cost $nd$ multiplications. These methods do not exploit sparsity of $G$ and the locality of computing a single component. 

There are nearly linear time\footnote{$O(m \log^c n \log \epsilon^{-1})$, where $m$ is the number of nonzero entries in $A$, and $c \in \Reals_+$ is a fixed constant.} approximation algorithms for sparse and symmetric diagonally dominant matrices $A$ (i.e. graph Laplacians), however they involve global structures over the graph, such as graph sparsifiers or spanning trees, and the goal is to estimate the entire solution vector \cite{SpielmanTeng06, KoutisMillerPeng11, Kelner13, Vishnoi13}. 

\medskip {\bf Asynchronous distributed algorithms.} In their seminal work, Bertsekas and Tsitsiklis \cite{BertsekasTsitsiklis89} analyzed the asynchronous implementation of stationary linear iterative methods for solving for the full vector $x$, where they assign each of $n$ processor to compute updates corresponding to a specific coordinate. They use a different computation model involving a network of distributed processors computing simultaneously, whereas our model involves a shared global memory through a distributed file system (DFS) and variable number of processors computing in parallel. The cost of our algorithm is considered in terms of computational resources consumed, i.e., the number of tasks and DFS accesses, whereas they consider the number of parallel computations until convergence, where each of the $n$ processors are computing at every time step. Our algorithm relies on residual based updates, maintaining an invariant that allows us to precisely characterize the error as a function of the residual vector. These differences lead to very different proof techniques for proving both eventual convergence as well as convergence rate bounds. 

There has also been work on distributed and asynchronous algorithms from an optimization standpoint \cite{LiuMouMorse13, MouLiuMorse15}. Minimizing the objective function $\|A x - b\|_2^2$ can be written as a distributed optimization task, where each computational node $u \in [n]$ aims to minimize $(e_u^T Ax - b_u)^2$ while seeking consensus such that all nodes converge to the same solution vector $x$. Again these algorithms focus on the global computation task of the full solution vector $x$ rather than estimating a single component. 

\medskip {\bf Local algorithms.} Methods for computing a single component can be categorized into either Monte Carlo methods which sample random walks, or deterministic iterative methods. The Ulam von Neumann algorithm is a Monte Carlo method which obtains an estimate for a single component $x_i$ by sampling random walks starting at the vertex $i$. It interprets the Neumann series representation of the solution $x$ as a sum over weighted walks on $\mathcal{G}(G)$, and obtains an estimate by sampling random walks starting from vertex $i$ over $\mathcal{G}(G)$ and appropriately reweighting to obtain an unbiased estimator \cite{Forsythe50, Wasow52, Curtiss54, DimovMaireSellier15}. The challenge is to control the variance of this estimator. The classic choice for the sampling matrix requires $\|G\|_{\infty} < 1$, though there are modifications which propose other sampling matrices or use correlated sampling to reduce the variance \cite{Halton70, Halton94}. The scope of this algorithm is limited, as Ji, Mascagni, and Li proved that there is a class of matrices such that $\rho(G) < 1$, $\|G\|_{\infty} > 1$, and there does not exist any sampling matrix such that the variance of the corresponding estimator is finite \cite{JiMascagniLi13}. In contrast, our algorithm exploits the sparsity of $G$ and provides a convergent solution when $\rho(G) < 1$ and convergence rates when $\|G\|_2 < 1$. Single component approximation of the leading eigenvector for a stochastic matrix has been studied using Monte Carlo random walk sampling methods \cite{Lee13}.

\cite{GleichKloster15} propose an iterative method for approximating a single column of the matrix exponential, which can also be written as a series of matrix powers, $\exp(G) = \sum_{k=0}^{\infty} \frac{1}{k!} G^k$, similar to the Neumann series in \eqref{eq:neumann}. The algorithm essentially runs coordinate descent to compute the solution to a linear system of the form $A x = e_i$, where $A$ is constructed in such a way that the solution $x$ is an approximation for $\exp(G)$. They provide convergence guarantees for $\|G\|_1 \leq 1$ for the Gauss-Southwell and Gauss-Seidel iterations. This method has been independently studied for the specific setting of computing Pagerank, Andersen {\it et al.} proposed an iterative method which relies on the conditions that $G$ is a nonnegative scaled stochastic matrix, $z$ is entry-wise positive and bounded strictly away from zero, and the solution $x$ is a probability vector (i.e., consisting of nonnegative entries that sum to 1) \cite{Andersen07}. There has been subsequent follow up work which builds upon an earlier version of our paper to design bidirectional local algorithms that combine both iterative algorithms and Monte Carlo methods \cite{ShyamkumarBanerjeeLofgren16, LeeOzdaglarShah14}.

\medskip {\bf Relationships to our algorithm.} 
Our asynchronous algorithm is an iterative method and can be interpreted as using coordinate descent with randomized coordinate selection to solve for $(G^T)^{-1} e_i$, and then taking the inner product of the result with $z$ to obtain $x_i$. Our algorithm is different from the global algorithms as it specifically targets approximating a single component using local computations. It is also different from the Monte Carlo methods which tend to have high variance and thus slow convergence. It is most similar to the algorithms proposed by \cite{GleichKloster15, Andersen07}, however our algorithm has a different choice of termination conditions, and different rules for choosing a coordinate update order, utilizing probabilistic sampling. This not only requires very different analysis, but also allows for the algorithm to be implemented in an asynchronous distributed manner without coordination between tasks.

The model assumptions and analysis are also different, as \cite{GleichKloster15, Andersen07} focus on stochastic matrices. The analysis of \cite{Andersen07} proves a linear decrease in the error, yet we prove that the second moment of our error contracts by a time dependent factor in each iteration, and thus our algorithm converges to the correct solution with a tighter convergence rate. We provide analysis of convergence considering the sparsity pattern of the matrix, and we show that any arbitrary coordinate selection rule converges as long as each coordinate is updated infinitely often. In contrast \cite{GleichKloster15, Andersen07} only guarantee convergence for specific coordinate update orders.

The use of randomization in subsampling matrices as part of a subroutine in iterative methods has previously been used in the context of other global matrix algorithms, such as the randomized Kaczmarz method and stochastic iterative projection \cite{StrohmerVershynin09, SabelfeldLoschina10, Sabelfeld11, Sabelfeld09, WangBertsekas13}. The randomized Kaczmarz method is used in the context of solving overdetermined systems of equations, subsampling rows to reduce the dimension of the computation matrix in each iteration. Stochastic iterative methods involve sampling a sparse approximation of matrix $G$ to reduce the computation in each iteration while maintaining convergence.


\section{Distributed Computation Model}\label{sec:model}

In the modern world of large scale computation, as the requirement for computational resources and memory storage increases, distributed cloud computing systems have become the norm for computation that involves handling large amounts of data. Since the computing power and memory of any single processor is limited, large distributed file systems (DFS), e.g. Hadoop-Distributed-File-System (HDFS), collect together many storage disks with a master node which handles I/O requests, allowing clients to access the information in the distributed file system in a similar way of accessing files from the local disk. An algorithm can be parallelized by separating it into small tasks that can each be computed by a single processor through accessing the DFS. The access time of I/O requests to the distributed file system is much longer than accessing files in a processor's own local memory, so we would like to minimize the number of DFS accesses in addition to the computing resources consumed, i.e. total number of tasks performed.

\begin{figure}
\centering
\includegraphics[width=0.5\textwidth]{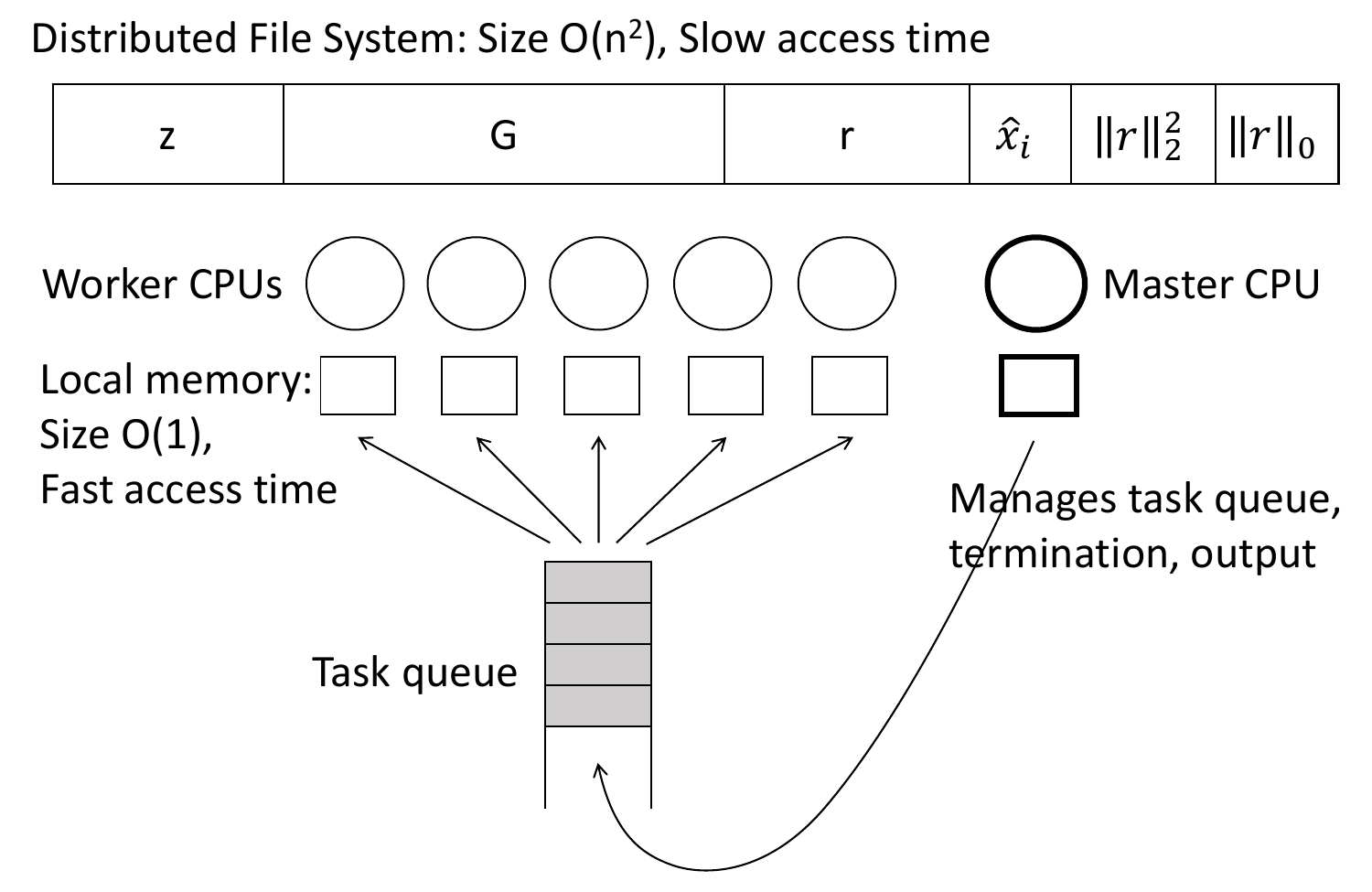}
\caption{Model of Parallel Computation with Distributed File System.}
\label{fig:comp_model}
\end{figure}

In this paper, we will assume the computation model as described in Figure \ref{fig:comp_model}. There is a large distributed file system, which all the processors have access to. There is a collection of processors (CPUs) each with a small fixed size local memory. One CPU is designated the master CPU, and it manages the task queue as well as the termination and output of the algorithm. The remaining processors are designated worker CPUs, and they perform tasks assigned to them from the task queue. The cost will be counted in terms of the amount of computing resources that the entire computation consumes, e.g. the number of tasks performed, DFS accesses per task, and storage used in the DFS. In many cloud computation systems, the computing resources are shared across many jobs that are running on the cloud, therefore, the task queue may include tasks corresponding to unrelated jobs as well. 

\section{Algorithm Intuition}

Given a vector $z$ and matrix $G$ such that $\rho(G) < 1$, our goal is to approximate the $i^{\text{th}}$ component of the solution vector $x$ to $x = Gx + z$. Classic stationary linear iterative methods use updates of the form $x^{(t+1)} = G x^{(t)} + z$ to iteratively approximate leading terms of the Neumann series. The matrix-vector multiplication can be performed in a distributed manner by splitting it into update tasks, where each task updates a single coordinate $u$ according to
\[x_u^{(t+1)} \leftarrow \textstyle\sum_{v \in N_u} G_{uv} x_v^{(t)} + z_u\]
for some $u \in \{1,2, \dots n\}$. These tasks are added to the task queue and assigned to different processors to compute. Since $z$ can be fully dense, the vectors $x^{(t)}$ will be at least as dense as $z$, thus computing $x^{(t+1)}$ from $x^{(t)}$ involves $n$ individual coordinate update tasks. In our problem, since we are specifically interested in a single component $i$, we instead define a residual-based update method which maintains sparsity of the intermediate vector involved in the computation. We will first present a synchronous distributed version of the algorithm. In section \ref{sec:alg_det}, we will present the asynchronous distributed implementation of the algorithm, and argue that even when the updates are performed asynchronously, the algorithm still converges to the correct solution. Both implementations require at most $O(n + |\mathcal{E}|)$ space in the DFS to store the matrix $G$, vector $z$, and any intermediate values involved in the computation.

According to the Neumann series representation of $x$, 
\begin{align}
x_i = e_i^T \textstyle\sum_{k=0}^{\infty} G^k z = z^T \textstyle\sum_{k=0}^{\infty} (G^T)^k e_i.
\end{align}
Consider defining the residual vector at iteration $k$ as $r^{(k)} = (G^T)^k e_i$. Observe that the sparsity pattern of $(G^T)^k e_i$ is given by $N_i(k)$, the set of vertices for which there is a path of length $k$ from vertex $i$. We can rewrite $x_i$ as a function of the residual vectors
\[x_i = z^T \textstyle\sum_{k=0}^{t-1} r^{(k)} + x^T r^{(t)}.\]
Let $\hat{x}_i^{(t)}$ denote our estimate of $x_i$ at iteration $t$. We can iteratively approximate $x_i$ with the low order terms of the Neumann series using the following updates:
\begin{align}
\hat{x}_i^{(t+1)} &\leftarrow \hat{x}_i^{(t)} + z^T r^{(t)}, \\
r^{(t+1)} &\leftarrow G^T r^{(t)},
\end{align}
and initializing with $\hat{x}_i^{(0)} = 0$, and $r^{(0)} = e_i$. Since the sparsity of $r^{(t)}$ is at most the size of the $t$-radius neighborhood of vertex $i$, denoted $|N_i(t)|$, the computation involved in one iteration can be split into $|N_i(t)|$ single coordinate update tasks, corresponding to updating each coordinate $u \in N_i(t)$. A task updating coordinate $u$ executes the following steps:
\begin{enumerate}
\item ADD $G_{uu} r_u^{(t)}$ to $r_u^{(t+1)}$,
\item ADD $z_u r_u^{(t)}$ to $\hat{x}^{(t+1)}_i$,
\item For each $v \in N_u$, ADD $G_{uv} r_u^{(t)}$ to $r_u^{(t+1)}$.
\end{enumerate}
Each update task uses at most O($|N_u|$) DFS accesses, and does not require more than constant space in the local memory. The processor can only store the value of $r_u^{(t)}$, and sequentially access and compute $G_{uv} r_u^{(t)}$ for $v \in N_u$, requiring $|N_u|$ DFS accesses, but only O(1) memory for stored information across computations. We initialize the estimate and residual vectors for the $(t+1)^{\text{th}}$ iteration with $r^{(t+1)} = 0$ and $\hat{x}_i^{(t+1)} = \hat{x}_i^{(t)}$.

The processors still need to pay a synchronization cost due to coordinating the iterations of computation. This results in delays as tasks for a new iteration must wait until every update task from the previous iteration completes. Termination can be determined by imposing a condition on the residual vector which is checked after each iteration of computation, such as terminating when $\|r\|_2 < \epsilon$. In section \ref{sec:synch_conv}, we will prove convergence rate bounds for the synchronous implementation and discuss the gains the algorithm attains from a coordinate-based computation as opposed to computing the full vector.

\section{Asynchronous Updates} \label{sec:alg_det}

The method described above requires coordination amongst the tasks to track each iteration of the algorithm. This may cause unnecessary delays due to enforcing that the tasks must be completed in a specific order. In this section, we introduce an asynchronous implementation of the algorithm, in which the update tasks may be performed in arbitrary order, and we do not need to wait for previous tasks to complete before beginning to compute a new task. In section \ref{sec:async_conv}, we prove that the algorithm always converges, and we establish convergence rate bounds for different coordinate update rules.

\subsection{Individual Update Tasks}

In the asynchronous implementation, since we no longer keep track of any iterations of the algorithm, we will simply store a single instance of the residual vector $r$ in the DFS. When the different tasks update their corresponding coordinates, they will read and write their updates to the residual vector stored on the DFS. The algorithm is initialized in the same way with $r = e_i$ and $\hat{x}_i = 0$. A task to update residual coordinate $u$ involves three steps:
\begin{enumerate}
\item READ $r_u$, and SET $r_u$ to $G_{uu} r_u$,
\item ADD $r_u z_u$ to $\hat{x}_i$,
\item For each $v \in N_u$, ADD $G_{uv} r_u$ to $r_v$.
\end{enumerate}
The value of $r_u$ used in steps 2 and 3 is the original value read from the DFS in step 1. For each task, the worker processor makes O($|N_u|$) DFS accesses. The computation is the same as individual tasks in the synchronous implementation, except without keeping track of the residual vector across distinct iterations. Thus, even when the sequence of coordinate updates is the same, the residual vector in the asynchronous implementation will evolve differently. If $v \in N_u$, and $v$ is updated after $u$, when the asynchronous implementation of the algorithm updates coordinate $v$, it will use the previously updated value in which the task corresponding to coordinate $u$ added $G_{vu} r_u$ to the value of $r_v$. In section \ref{sec:walks}, we introduce an interpretation of the algorithm as summing weighted walks in the graph. The synchronous implementation sums the walk in a breadth first manner, i.e. all walks of length $t$ are summed in the $t^{\text{th}}$ iteration, whereas the asynchronous implementation may sum walks of different lengths in a single update.

For the purposes of analyzing the convergence rate bounds, we consider that the three steps involved in a single update task are performed together as a single unit of computation, i.e., that the different steps involved in a single update task are executed together, and do not interleave with other tasks. We let $\hat{x}_i^{(t)}$ denote the estimate after $t$ update tasks have completed, and we let $r^{(t)}$ denote the residual vector after $t$ update tasks have completed. This property can be enforced through read and write locks, which would prevent another task from simultaneously changing the value of $r_u$ while a particular task is in the middle of computation involving $r_u$. This allows us to clearly track the value of the residual vector after each update task, lending to convergence rate bounds. 

However, we will be able to prove asymptotic convergence with much weaker conditions, in which only step 1 of the update task needs to be considered a single unit executed together. Since addition operations are exchangeable, the correctness of the algorithm still holds even when the addition operations in step 2 and 3 of the update task may interleave with other operations on the data from other tasks. Step 1 needs to be executed together because we need to make sure that another task does not add value to $r_u$ in between the time that we first read $r_u$ and write $G_{uu} r_u$, since we would then accidentally override the added value. Alternatively, we would not want another task to read the same value of $r_u$ and begin repeating the same update that we have already begun. We will show in Lemma \ref{lemma:asynch_invariant} that the invariant $\hat{x}_i - x_i = r^T x$ holds before and after any update task.

\subsection{Coordinate Update Rule}

In the asynchronous implementation, we are given more flexibility to choose the order in which the coordinates are updated. We could update in the same order as the synchronous implementation, in which we round robin update coordinates according to a breadth first traversal over the graph (yet allowing for loops and repeated visits to vertices), i.e., updating first all coordinates in $N_i(1)$, followed by $N_i(2)$. Similarly we can iterate round robin updates for all coordinates with nonzero residual vector values. This can be coordinated by designating one processor as the ``master'', whose job is to add tasks to the the task queue.

In settings where we would like to elimination coordination between tasks from a master processor, we can use randomization to generate the tasks or coordinate update order. To approximate the round robin procedure, we could probabilistically choose the next update coordinate by sampling uniformly randomly from all coordinates with nonzero residual values, which we term the `uniform censored sampling' procedure. As each processor finishes a task, it can generate the next task by sampling a new update coordinate. This can be facilitated by storing the value of $\|r\|_0$ as well as a list of coordinates with nonzero valued residuals, and the update tasks can easily be modified to maintain the value of $\|r\|_0$ and list of relevant coordinates.

As our algorithm is derived from residual based updates, and the estimation error is given by $r^T x$, this suggests that we may make more progress if we focus on updating coordinates with large residual values. For example, we can choose to always update the coordinate with the largest residual value. This can be implemented by maintaining a priority queue with the residual values. We could also sample a coordinate probabilistically proportional to some function of the residual, e.g., proportional to $|r_u|$, or $r_u^2$. This may be more difficult to implement without iterating through the residual vector to generate each sample, though it is still possible to implement in our distributed computation model with a larger number of DFS accesses.

\subsection{Termination}

The termination conditions can be chosen depending on the desired accuracy and the value of the residual vector. The error is given by $r^T x$, but since we do not know the value of $x$, we can design the termination condition as a function of $r$. For example, terminating when $\|r\|_2 < \epsilon$ results in an additive error bound of at most $\epsilon \|x\|_2$. The individual update tasks can be modified to additionally keep track of $\|r\|_2^2$, $\|r\|_1$, or $\|r\|_{\infty}$ without incurring much overhead, since these quantities are additive across coordinates, and each update task changes at most $d+1$ coordinates of $r$.

We are motivated by network analysis settings in high dimension, such as computing Pagerank or Bonacich centrality when $n$ is large. As $n$ grows to infinity for some large graph, $\|x\|_2$ is in fact normalized, bounded, and doesn't scale with $n$ for these three example network centralities. Most of the mass is contained in a few components, implying that an additive error bound of $\epsilon \|x\|_2$ approximately guarantees a multiplicative error for large weight components, and an additive error for small weight components. Therefore, we will present many of our results assuming the algorithm uses  a termination condition of $\|r\|_2 < \epsilon$.

\section{Synchronous Analysis} \label{sec:synch_conv}

In order to compare the convergence rate bounds for the asynchronous implementation, we first analyze the synchronous implementation. We will count the number of tasks and multiplications that the synchronous implementation uses. This analysis will help us to appreciate and identify the gains the algorithm makes due to sparsity and local computation.

\begin{theorem}\label{thm:synch}
If $\rho(G) < 1$, the synchronous implementation of the algorithm converges, and estimation error decays as
\[|\hat{x}_i^{(t)} - x_i| = r^{(t)} x \leq \|G\|_2^t \|x\|_2.\]
The total number of update tasks the algorithm performs in $t$ iterations is
\[O\left(\textstyle\sum_{k=0}^{t-1} |N_i(k)|\right) = O\left(\min\left(d^t, nt\right)\right),\]
where $N_i(k)$ is the set of vertices which are within a $k$-radius neighborhood of vertex $i$, and $d = \max_u |N_u|$. The number of DFS accesses per task is bounded above by $d$.
\end{theorem}

Corollary \ref{corr:sync} follows from the proof of Theorem \ref{thm:synch}, and highlights that if the graph is sparse, or the size of the neighborhood grows slowly, then the complexity of the algorithm can scale much better than computing the entire solution vector, which would cost $O(n\ln(\epsilon)/\ln(\|G\|_2))$ update tasks.

\begin{corollary} \label{corr:sync}
If we terminate the algorithm when \\ $\|r^{(t)}\|_2 < \epsilon$, then 
$|\hat{x}_i - x_i| \leq \epsilon \|x\|_2$, and the total number of update tasks performed is bounded by
\[O\left(\min\left(\epsilon^{\ln(d)/\ln(\|G\|_2)}, \tfrac{n\ln(\epsilon)}{\ln(\|G\|_2)}\right)\right).\]
The number of DFS accesses per task is bounded above by $d = \max_u |N_u|$.
\end{corollary}

\begin{proof}[Proof of Theorem \ref{thm:synch}]
The initial vectors and update rules are chosen to satisfy the invariant that for all $t$, $x_i = \hat{x}_i^{(t)} + x^T r^{(t)}$. The error in the estimate at iteration $t$ is given by $x^T r^{(t)} = x^T (G^{t})^T e_i$. When $\rho(G) < 1$, the error converges to zero, and thus the algorithm converges. It follows that the error is bounded by
\begin{align}
|\hat{x}_i - x_i | &= | r^{(t)T} x| \leq \|r^{(t)}\|_2 \|x\|_2.
\end{align}
When the algorithm terminates at $\|r^{(t)}\|_2 \leq \epsilon$, the error is bounded by $\epsilon \|x\|_2$, and after $t$ iterations, the error is bounded by $\|(G^T)^t e_i\|_2 \|x\|_2 \leq \|G\|_2^t \|x\|_2$.
Since each row of $G$ has at most $d$ nonzero entries, $\|r^{(t)}\|_0 \leq \min(d^t,n)$. The number of coordinate update tasks in each iteration is at most
$\|r^{(t)}\|_0 \leq |N_i(k)| \leq \min\left(d^t, nt\right).$
Therefore, we can upper bound the total number of tasks in $t$ iterations by by
\[O\left(\textstyle\sum_{k=0}^{t-1} |N_i(k)|\right) = O\left(\min\left(d^{t}, nt\right)\right).\]
Since $\|r^{(t)}\|_2$ decays as $\|G\|_2^t$, the algorithm terminates at $\|r\|_2 < \epsilon$ within at most $\ln(\epsilon)/\ln(\|G\|_2)$ iterations, upper bounding the tasks performed by
\[O\left(\min\left(\epsilon^{\ln(d)/\ln(\|G\|_2)}, \tfrac{n\ln(\epsilon)}{\ln(\|G\|_2)}\right)\right).\]
\end{proof}

When $G$ is nonsymmetric, $\rho(G) < 1$, yet $\|G\|_2 \geq 1$, the algorithm still converges asymptotically, though our rate bounds no longer hold. We suspect that in this case, a similar convergence rate holds as a function of $\rho(G)$, due to Gelfand's spectral radius formula, which states that $\rho(M) = \lim_{k \rightarrow \infty} \|M^k\|_2^{1/k}$. The precise rate depends on the convergence of this limit.

The right hand expression in the theorem comes from bounding the number of coordinate updates in each iteration by $n$, which holds even in the nonsparse setting. This bound obtains the same result as standard linear iterative methods. The analysis of our algorithm highlights the improvement of our local algorithm over a general global vector computation. The number of tasks grows as $O(\textstyle\sum_{k=0}^{t-1} |N_i(k)|)$, which for some graphs may be significantly less than $O(nt)$. For a bounded degree graph,
$\textstyle\sum_{k=0}^t |N_i(k)| = O(d^t),$
which may be much less than $O(nt)$ in the case when $d$ is fixed, and $n$ is very large (recall that $t$ is on the order of $\ln(\epsilon)/\ln(\|G\|_2)$). The bounded degree condition is used in our analysis to cleanly bound $|N_i(t)|$, however our results naturally extend to other graphs given bounds on $|N_i(t)|$. For power law graphs, we can use a bound on the growth of the local neighborhood size for average vertices to obtain non-trivial convergence rate results. For graphs in which the size of the neighborhood only grows polynomially, then the local algorithm would gain significant savings over the global algorithm. This results in  conditions under which our algorithm achieves an approximation for $x_i$ in constant time with respect to the size of the matrix for large $n$, e.g. $d = O(1)$ and $-1/\ln(\|G\|_2) = O(1)$.

We can visualize the algorithm in terms of computation over $\mathcal{G}(G)$. Multiplying $r^{(t)}$ by $G^T$ corresponds to a message passing operation from each of the nonzero coordinates of $r^{(t)}$ along their adjacent edges in the graph. The sparsity of $r^{(t)}$ grows according to the set of length $t$ walks over the graph that originate from vertex $i$. The termination condition guarantees that the algorithm only involves vertices that are within distance $\ln(\epsilon)/\ln(\|G\|_2)$ from the vertex $i$. We define the matrix $G_{N_i(t)}$ such that $G_{N_i(t)}(a,b) = G(a,b)$ if $(a,b) \in N_i(t) \times N_i(t)$, and is zero otherwise. It follows that 
\begin{align}
\|r^{(t)}\|_2 &= \|e_i^T G^t\|_2 = \|e_i^T \textstyle\prod_{k=1}^t G_{N_i(k)}\|_2 \\
&= \|e_i^T G_{N_i(t)}^t\|_2 \leq \|G_{N_i(t)}\|_2^t.
\end{align}
It is possible that for some choices of $i$ and $t$, $\|G_{N_i(t)}\|_2 < \|G\|_2$, in which case the algorithm would converge more quickly as a function of the local neighborhood. If $G$ corresponds to a scaled adjacency matrix of an unweighted undirected graph, then it is known that
\begin{align}
\max\left(d_{\text{average}},\sqrt{d_{\max}}\right) \leq \rho(G) \leq d_{\max}.
\end{align}
In this case, we would only expect $\|G_{N_i(t)}\|_2$ to be smaller than $\|G\|_2$ if the local degree distribution of the neighborhood around vertex $i$ is different from the global degree distribution.

\section{Asynchronous Analysis} \label{sec:async_conv}

It is not as straightforward to analyze the asynchronous implementation of the algorithm, since we can no longer write the residual vector as a simple expression of $G$ and the iteration number. However, we can show that each coordinate update task preserves an invariant which relates the estimate $\hat{x}_i$ and residual $r$ to the true solution $x_i$.

\begin{lemma}[Invariant] \label{lemma:asynch_invariant}
The update tasks in the asynchronous implementation maintain the invariant that for all $t$, $x_i = \hat{x}_i + r^T x$.
\end{lemma}

\begin{proof}[Proof of Lemma \ref{lemma:asynch_invariant}]
Recall that $x = z + G x$. We prove that the invariant holds by using induction. First verify that before any computation has begun, the invariant is satisfied by the initialized values,
\[\hat{x}_i + r^T x = 0 + e_i^T x = x_i.\]
Let $r^{old}$ denote the residual vector before an update task, and let $r^{new}$ denote the residual vector after an update task. Then a single update task corresponds to the following steps:
\begin{align*}
\hat{x}_i^{new} &= \hat{x}_i^{old} + r^{old}_u z_u, \\
r^{new}_u &= G_{uu} r^{old}_u, \\
r^{new}_v &= r^{old}_v + G_{uv} r^{old}_u, \forall v \in N_u.
\end{align*}
Assuming that $x_i = \hat{x}_i^{old} + \langle r^{old} x \rangle$, it follows that
\begin{align*}
&\hat{x}_i^{new} + \langle r^{new} x \rangle - \hat{x}_i^{old} - \langle r^{old} x \rangle \\
&= r^{old}_u z_u + x_u (G_{uu} - 1) r^{old}_u + \textstyle\sum_{v \in N_u} x_v G_{uv} r^{old}_u, \\
&= e_u^T (z + G x - x) r^{old}_u = 0.
\end{align*}
 \end{proof}

It follows from Lemma \ref{lemma:asynch_invariant} that we can choose termination conditions based upon the value of the residual vector which would directly lead to upper bounds on the estimation error. For example, if $\|r\|_2 \leq \epsilon$, then $|\hat{x}_i - x_i| \leq \epsilon \|x\|_2$. The proofs for Theorems presented in the subsequent sections for the asynchronous algorithm can be found in sections \ref{sec:pf1}, \ref{sec:pf2}, and \ref{sec:pf3}.

\subsection{Counting Weighted Walks} \label{sec:walks}

Alternatively, we can take the perspective that the algorithm is computing $x_i$ by collecting a sum of weighted walks over the graph $\mathcal{G}(G)$ beginning at vertex $i$. The estimate $\hat{x}_i$ corresponds to the sum of all weighted walks which are already ``counted'', and the residual vector represents all yet uncounted walks. As long as step 1 of the coordinate update task is atomic, we can ensure that every walk is accounted for exactly once, either in $\hat{x}_i$, or in the residual vector. Let $\tilde{G}$ denote the matrix where $\tilde{G}_{ij} = |G_{ij}|$. Theorem \ref{thm:asynch_asymptotic} uses the perspective of counting weighted walks to show that as long as $\rho(\tilde{G}) < 1$, the algorithm converges to $x_i$ as long as each coordinate is chosen infinitely often, regardless of the sequence in which the updates are performed.

\begin{theorem}\label{thm:asynch_asymptotic}
If $\rho(\tilde{G}) < 1$, the estimate $\hat{x}_i$ from the asynchronous implementation of our algorithm converges to $x_i$ for any sequence of coordinate updates, as long as each coordinate is updated infinitely often.
\end{theorem}

The solution $x_i$ can be expressed as a weighted sum over all walks over the graph $\mathcal{G}(G)$ beginning at vertex $i$, where a walk beginning at vertex $i$ and ending at vertex $j$ has weight $\textstyle\prod_{e \in \text{walk}} G_e z_j$. The updates ensure that we never double count a walk, and all uncounted walks are included in the residual vector $r$. For any $l$, there is a finite time $S_l$ after which all random walks of length less than or equal to $l$ have been counted and included into $\hat{x}_i$. This allows us to upper bound $|x_i - \hat{x}_i|$ as a function of $\tilde{G}^{l+1}$, which converges to zero when $\rho(\tilde{G}) < 1$. If $\rho(\tilde{G}) \geq 1$, then the original Neumann series stated in \eqref{eq:neumann} is only conditionally convergent. By the Reimann series theorem, the terms can be rearranged in such a way that the new series diverges, and rearranging the terms in the series corresponds to updating the coordinates in different orders, e.g., depth first as opposed to breadth first. This is the same conditions for asymptotic convergence as provided for the asynchronous linear iterative updates, which is also shown to be tight \cite{BertsekasTsitsiklis89}. This theorem and proof can be extended to show that the algorithm converges asymptotically even given communication delays, as long as the messages reach their destination in finite time.

In fact our proof for the asymptotic convergence translates directly into a convergence rate bound as well.

\begin{theorem} \label{thm:sl_bound}
Suppose the asynchronous implementation of our algorithm used the coordinate update sequence $(u_0, u_1, u_2, u_3, \dots )$, where each coordinate updates infinitely often. Define $S_r$ as the time after which the estimate vector has counted and included all walks of length up to $r$:
\[S_l = \min\{t \geq S_{l-1}: N_i(l) \subset \{u_{S_{l-1}}, u_{S_{l-1} + 1}, \dots u_{t-1}\}\}.\]
Then the estimation error of the the algorithm after $S_l$ updates is bounded by
\[\left|x_i - \hat{x}_i^{(S_l)}\right| \leq x^T (\tilde{G}^T)^{l+1} e_i \leq \|\tilde{G}\|_2^{l+1} \|x\|_2.\]
\end{theorem}

Based upon the update sequence we can compute bounds on $S_l$, or the time after which all walks of length $l$ have definitely been counted. We can analyze the basic coordinate update rule which follows the same pattern as the synchronous implementation, in which we update according to the neighborhoods of $i$. The update sequence would be given by $(N_i(0), N_i(1), N_i(2), \dots)$. Although the update order may be the same as the synchronous algorithm, the computation is not the same due to the accumulation of the residual vector across update tasks. It follows that due to the update order, $S_l = \textstyle\sum_{k=0}^l |N_i(k)|.$
If the graph is bounded degree with max degree $d$, such that $|N_i(k)| \leq d^k$, then $S_l \leq d^{l+1}$, resulting in the following corollary.

\begin{corollary}
Suppose the asynchronous implementation of our algorithm used the coordinate update sequence $(N_i(0), N_i(1), N_i(2), \dots)$. Then the estimation error of the algorithm after $d^{t+1}$ update tasks is bounded by
\[\left|x_i - \hat{x}_i\right| \leq \|\tilde{G}\|_2^{l+1} \|x\|_2.\]
It follows that the error is less than $\epsilon \|x\|_2$ for $t \geq d^{\ln(\epsilon)/\ln(\|\tilde{G}\|_2)}$.
\end{corollary}

This matches the convergence rate bound for the synchronous algorithm when $G$ is nonnegative, which is reasonable for some applications in which $G$ is derived from network data.
It also follows directly from Theorem \ref{thm:sl_bound} that if every coordinate updates at least once within every $B$ timesteps, then the error decays with rate $\|\tilde{G}\|_2^{t/B}$, which is comparable to the bounded delay model and analysis for the asynchronous linear iterative algorithm \cite{BertsekasTsitsiklis89}.

\begin{corollary}
Suppose the asynchronous implementation of our algorithm used a coordinate update sequence in which $S_l \leq lB$ for some $B > 0$. Then the estimation error of the the algorithm after $lB$ updates is bounded by
\[\left|x_i - \hat{x}_i\right| \leq \|\tilde{G}\|_2^{l+1} \|x\|_2.\]
It follows that the error is less than $\epsilon \|x\|_2$ for $t \geq B \ln(\epsilon)/\ln(\|\tilde{G}\|_2)$.
\end{corollary}

\subsection{Probabilistic Update Order} \label{sec:prob_conv}

When the coordinates are sampled probabilistically, we can no longer guarantee that a certain set of coordinates are updated within a fixed interval. In this section, we instead provide a probabilistic analysis of the error by analyzing the evolution of the 2-norm of the residual vector in expectation. We will assume that each coordinate update task is atomic, such that if the sequence of coordinate updates is given by $(u_0, u_1, u_2, \dots)$, the residual vector after $t$ updates will be equivalent to the following computation:
\[r = \left(\textstyle\prod_{s=0}^{t-1} (I - e_{u_s} e_{u_s}^T (I - G))\right)^T e_i.\]
The precise expression depends on the detailed order of updates, and thus the convergence rate may depend upon the rule that the algorithm chooses to determine the order of updating coordinates. 

We provide an analysis for `uniform censored sampling', in which coordinates with nonzero valued current residuals are chosen with equal probability, according to
\begin{align}
\mathcal{P}(u) = \tfrac{\mathbb{I}\left(r_u^{(t)} \neq 0\right)}{\|r^{(t)}\|_0}, \label{eq:sampleDistPhat}
\end{align}
where $r^{(t)}$ denotes the current residual after $t$ updates. We have suppressed the dependence of $\mathcal{P}$ on $r^{(t)}$ for simpler notation. Since the distribution $\mathcal{P}$ chooses uniformly among the nonzero coordinates of $r^{(t)}$, in expectation the update step corresponds to multiplying a scaled version of matrix $G$ to vector $r^{(t)}$. We can prove that in addition $\|r^{(t)}\|_2$ contracts with high probability due to the choice of distribution $\mathcal{P}$. With high probability, the number of multiplications the asynchronous algorithm uses is bounded by a similar expression as the bound given for the synchronous algorithm. 

\begin{theorem}\label{thm:asynch}
If $\|G\|_2 <1$, with probability 1, the asynchronous implementation which updates coordinates according to $\mathcal{P}$ eventually terminates at $\|r\|_2 < \epsilon$ and produces an estimate $\hat{x}_i$ such that $|\hat{x}_i - x_i| \leq \epsilon \|x\|_2$. With probability greater than $1 - \delta$, the total number of update tasks bounded by
\[O\left(\min\left(\left(\epsilon \sqrt{\delta/2}\right)^{-d/(1-\|G\|_2)}, \tfrac{-n \ln (\epsilon \sqrt{\delta})}{1-\|G\|_2} \right)\right).\]
The number of DFS accesses per task is bounded above by $d = \max_u |N_u|$.
\end{theorem}

In order to prove this result, we first show that 
\begin{align}
\E_{\mathcal{P}}\left[\left.r^{(t+1)} \right| r^{(t)}\right] &= \left(I - \left(\tfrac{I - G^T}{\|r^{(t)}\|_0}\right) \right) r^{(t)}.
\end{align}
This implies that in expectation, the error contracts in each update task by at least $(1 - (1-\|G\|_2)/\min(td,n))$. We use this to prove an upper bound on the expected L2-norm of the residual vector after $t$ update tasks, and we apply Markov's inequality to prove that the algorithm terminates with high probability within a certain number of multiplications. There is additional technical detail in the formal proof, as it needs to handle the fact the $\|r^{(t)}\|_0$ is dependent on the full history of the previous iterations. We first analyze the algorithm for a modified distribution, where the scaling factor grows deterministically according to $\min(td, n)$ as opposed to $\|r^{(t)}\|_0$, and we use a coupling argument to show that the upper bound on the termination time of the algorithm using the modified distribution translates to the original distribution $\mathcal{P}$. This establishes an upper bound on the Lyapunov exponent for a product of random matrices drawn from a time-dependent distribution.

This bound grows exponentially in $d$, while the corresponding bound in Corollary \ref{corr:sync} grows only polynomially in $d$. The rate of convergence of the asynchronous variant is slower by a factor of $d/\ln(d)$ because the provable contraction of the error in each iteration is now spread out among the nonzero coordinates of the current iterate.

Our convergence rate bounds only apply when the algorithm samples uniformly among nonzero coordinates of the residual vector $r^{(t)}$, according to the distribution $\mathcal{P}$ defined by \ref{eq:sampleDistPhat}. However, this may not be the distribution which optimizes the convergence rate. Choosing a distribution which is not uniform amongst the nonzero coordinates is analogous to multiplying the residual vector by a reweighted matrix $\tilde{G}$ in expectation, where each row of $\tilde{G}$ corresponds to a row of $G$ weighted by the probability of choosing that row in the new distribution. This is challenging to analyze, as the weights for each row may be dependent upon the entire history of update tasks. Analysis would requires characterizing the Lyapunov exponent of a product of random matrices, where each matrix is sampled from a different distribution, dependent upon the entire history, making it difficult to directly analyze the algorithm in expectation. Under stronger conditions (e.g. only nonnegative entries in $G, z$), we can show monotonic decrease in the error, and hence bound convergence rates for other sampling distributions with standard techniques.

\section{Simulations}

We implemented our algorithm on synthetic data to illustrate the convergence rates of different coordinate update order rules. In each of these examples, we sample a random graph and let the matrix $A$ denote the edge adjacency matrix of the graph. A scalar $\alpha$ is chosen small enough such that $\|\alpha A\|_2 \leq \|\alpha A\|_{\infty} \leq 0.9$. The goal is to compute the Bonacich centrality of a fixed vertex in the graph, given by the component of the solution vector $x$ to $x = \alpha A x + {\bf 1}$. The first graph is sampled from a Erdos-Renyi model with 1000 vertices, each edge being present independently with probability 0.0276. The second graph is sampled from a configuration model with 500 vertices and a power law degree distribution, $\Prob(\text{degree }d) \propto d^{-1.5}$.

We implement the synchronous implementation of our algorithm, and the asynchronous implementation algorithm with five choices of update rules. Round robin refers to the update rule which follows the expanding neighborhoods of vertex $i$, i.e., updating according to the sequence $(N_i(0), N_i(1), N_i(2), \dots)$. We implement the uniform censored sampling rule, which samples uniformly amongst nonzero valued residual coordinates. We explore sampling rules which depend on the value of the residuals, choosing the coordinate proportional to $|r_u|$, $r_u^2$, or chosen as $\text{argmax}_u |r_u|$. We compare with the standard linear iterative method, which uses updates of the form $x_{t+1} = \alpha A x_t + z$ to recursively approximate leading terms of the Neumann series, computing the full solution vector $x$. It is a global algorithm, as each iteration may involve multiplying a matrix with a dense vector, and thus simulations show it performs more poorly than our local algorithms. This insight should hold for other global algorithms as well, and the discrepancy will increase as the size of the graph increases.

\begin{figure}
\centering
	\subfigure[Erdos-Renyi network.]{
		\includegraphics[width=0.4\textwidth]{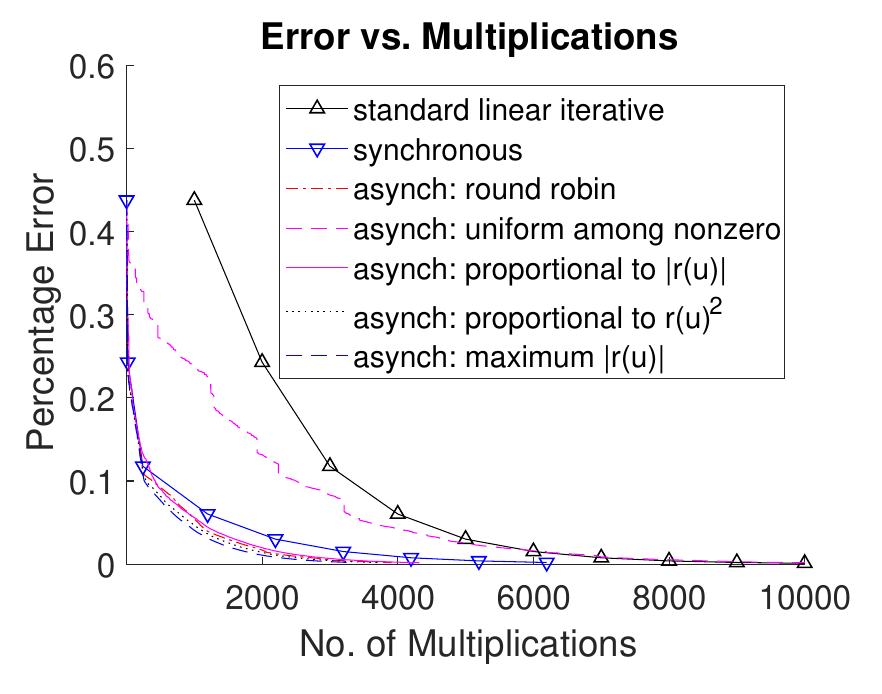}
		\label{fig:compare_P_rates_ER}
	}
	\subfigure[Power law degree network.]{
		\includegraphics[width=0.4\textwidth]{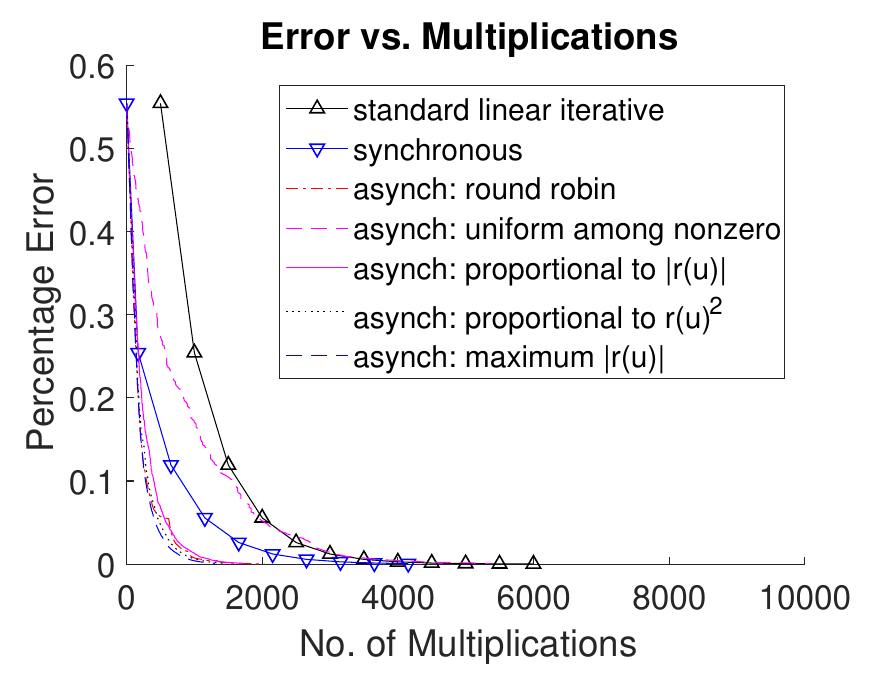}
		\label{fig:compare_P_rates_PL}
	}
	\caption{Comparing different implementations for computing Bonacich centrality of a vertex in a network.}
\end{figure}

Figures \ref{fig:compare_P_rates_ER} and \ref{fig:compare_P_rates_PL} show the percentage error of each algorithm with respect to the number of multiplications the algorithm has computed, for the Erdos Renyi graph and the power law graph respectively.
Our simulations indicate that choosing coordinates with large values of $|r_u|$ improves the convergence of the algorithm. The algorithm which always chooses the largest coordinate to update seems to perform the best. This is consistent with our intuition, as the residual vector $r$ is directly related to the estimation error. By updating coordinates with large values of $r_u$, we make more progress in reducing the error. Establishing theoretical analysis of this observation remains a challenging problem, as the sampling distribution for update task $t$ depends in a complex manner upon the full sample path of updates up to iteration $t$, as opposed to a simple scaling of the matrix as in uniform sampling. It is not obvious how to establish bounds for a product of random matrices drawn from a complex path dependent distribution.

We also observe that the our algorithm exhibits larger gains in the beginning of the algorithm, but the gains becomes less significant as the algorithm progresses. This could be a result of our algorithm exploiting the small size of the the local neighborhood in the beginning of the algorithm. As the size of the neighborhood grows in include all coordinates, our algorithm no longer enjoys sparse residual vectors, and thus the computational savings slows down.

\begin{figure}
\centering
		\includegraphics[width=0.5\textwidth]{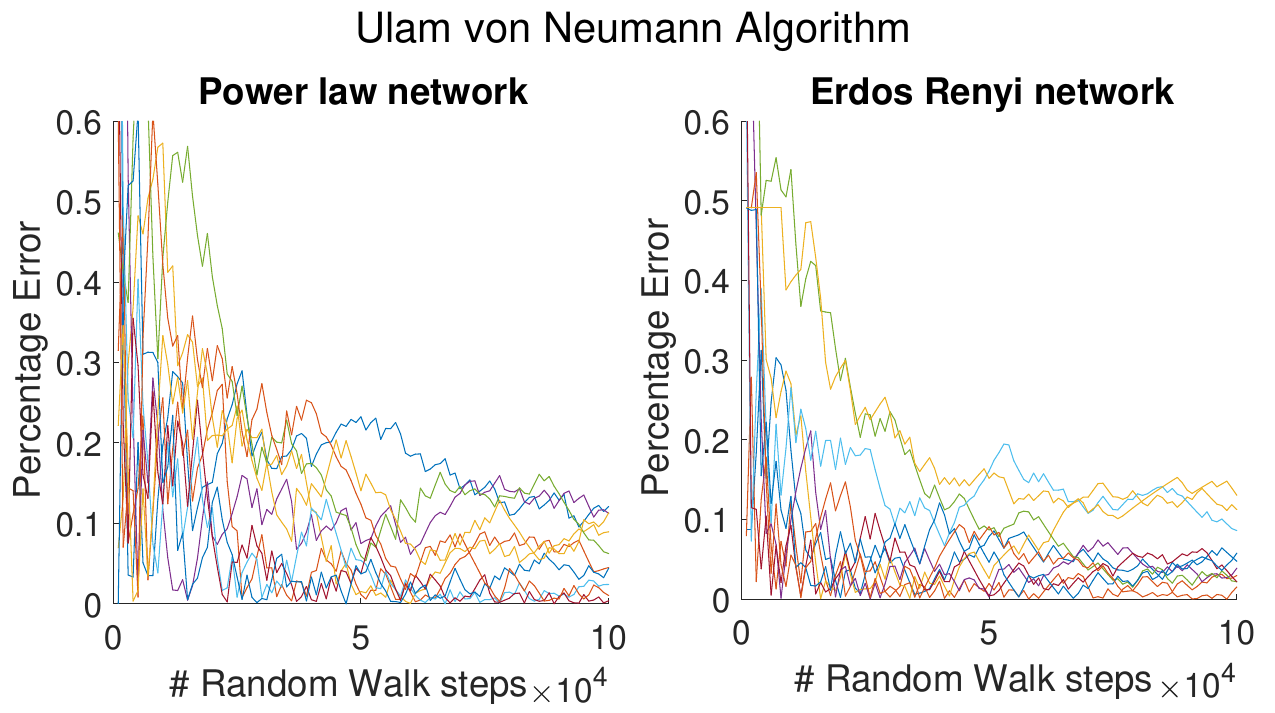}
		\label{fig:joint_UvN}
	\caption{Sample paths of 10 different instances of the Ulam von Neumann algorithm for computing Bonacich centrality of a vertex in a network.}
\end{figure}

We also provide results from using the Ulam-von-Neumann (UvN) Monte Carlo approach which samples random walks according to a transition probability matrix designed as $P_{uv} = \alpha A_{uv}$. All walks begin at the target vertex $i$, and the probability of the random walk terminating at vertex $u$ is $1 - \alpha \sum_v A_{uv}$. A random walk which terminates at vertex $u$ is given the weight $\frac{1}{1-\alpha \sum_v A_{uv}}$, such that the expected weight of a random walk is equal to the desired Bonacich centrality $x_i = e_i^T \sum_{k=0}^{\infty} (\alpha A)^k \bOne$. The algorithm samples many random walks and averages the weights to approximate $x_i$. In Figure \ref{fig:joint_UvN}, the left and right plots each show 10 sample paths obtained by the UvN algorithm for the Erdos-Renyi and power law graphs respectively. Each curve represents one instance of the algorithm. Whereas the iterative algorithms converged by about $5000$ multiplications, the UvN algorithm has high variance and does not converge even after $100000$ random walk steps.

%

\section{Future Directions}

It is an open problem to investigate the optimal choice of the probability distribution for sampling coordinates used within the asynchronous method, as simulations indicate that some distributions exhibit faster convergence rates. This is related to recent work which investigates weighted sampling for the randomized Kaczmarz method, although their methods depend on weighting according to the matrix $G$ rather than the values in the intermediate vectors of the algorithm. As our algorithm is also related to coordinate descent, there is related work showing that coordinate descent converges more quickly when coordinates are chosen according to the Gauss-Southwell method (those with largest residuals first) rather than random selection \cite{NutiniSchmidtLaradjiFriedlanderKoepke15}.

We hope that our work will initiate studies of sparsity-preserving iterative methods in asynchronous and distributed settings from an algorithmic perspective. Our methods could be used as subroutines in other methods in which there is a need for local or asynchronous matrix computation. 
It is unclear whether our algorithm achieves an optimal convergence rate. There is a wealth of literature which studies acceleration or preconditioning techniques for classic linear system solvers, and it would be interesting to see whether these acceleration techniques could be used to speed up the convergence of our local algorithm.

\section{Proof of Theorems \ref{thm:asynch_asymptotic} and \ref{thm:sl_bound}} \label{sec:pf1}

\begin{proof}[Proof of Theorems \ref{thm:asynch_asymptotic} and \ref{thm:sl_bound}]
Let $\mathcal{W}_k(i)$ denote the set of length $k$ walks beginning from vertex $i$, i.e., a sequence of vertices ${\bf w} = (w_0, w_1, w_2, \dots w_k)$ such that $w_0 = i$ and $(w_s, w_{s+1}) \in E$ for all $s \in \{0, 1, \dots k-1\}$. Then 
\begin{align}
x_i &= e_i^T \textstyle\sum_{k=0}^{\infty} G^k z = \textstyle\sum_{k=0}^{\infty} \textstyle\sum_{{\bf w} \in \mathcal{W}_k(i)} \textstyle\prod_{s = 0}^{k-1} G_{w_s w_{s+1}} z_{w_k}, \nonumber \\
&= z_i + \textstyle\sum_{k=1}^{\infty} \textstyle\sum_{{\bf w} \in \mathcal{W}_k(i)} \textstyle\prod_{s = 0}^{k-1} G_{w_s w_{s+1}} z_{w_k}, \nonumber \\
&= z_i + \textstyle\sum_{j \in N_i} G_{i j} \textstyle\sum_{k=0}^{\infty} \textstyle\sum_{{\bf w} \in \mathcal{W}_k(j)} \textstyle\prod_{s = 0}^{k-1} G_{w_s w_{s+1}} z_{w_k}, \nonumber \\
&= z_i + \textstyle\sum_{j \in N_i} G_{i j} x_j. \label{eq:recursiveWalk}
\end{align}
This shows that due to the form of the weights over each walk, we can express the weighted sum of all walks whose first edge is $(i,j)$ by $G_{ij} x_j$. Similarly, this argument extends recursively to show that the weighted sum of all walks whose first $l+1$ vertices are given by $(v_0, v_1, v_2, \dots v_l)$ is equivalent to 
$\left(\textstyle\prod_{s=0}^{l-1} G_{v_s v_{s+1}}\right) x_{v_l},$
since $x_{v_l}$ captures the sum and weights of the remaining unfinished portion of the walks. In other words, the weighted sum of walks with a certain prefix is equal to the product of the weights from the prefix portion of the walk multiplied with the sum of the weights for all walks that could continue from the endpoint of the prefix walk. The value of the residual vector $r_u$ contains the product of the weights along the prefix portion of the walks that end at vertex $u$, thus explaining why $r^T x$ is equivalent to the weight of all yet uncounted walks.

We use this perspective to interpret the single coordinate updates in the algorithm to argue that there is conservation of computation, i.e., we never double count a walk, and all uncounted walks are included in the residual mass vector $r$. The update in our algorithm can be interpreted through \eqref{eq:recursiveWalk}, as the first term $z_u$ captures the walks that end at vertex $u$, and each term within the summation $G_{uw} x_w$ counts the walks which take the next edge $(u,w)$. Each time that a coordinate $u$ is updated, the step 
\begin{enumerate}
\item[2.] ADD $r_u z_u$ to $\hat{x}_i$,
\end{enumerate}
corresponds to counting the walks which end at vertex $u$, and whose weight is contained within $r_u$. We also need to count the walks which have the same prefix, but do not yet terminate at vertex $u$, which is captured by multiplying $r_u$ by each of its adjacent vertices $G_{uw}$ and adding that to $r_w$:
\begin{enumerate}
\item[1.] SET $r_u$ to $G_{uu} r_u$,
\item[3.] For each $v \in N_u$, ADD $G_{uv} r_u$ to $r_v$.
\end{enumerate}
We proceed to argue that for any $l$, there is a finite time $t$ after which all random walks of length less than or equal to $l$ have been counted and included into the estimate $\hat{x}_i$.

Given a coordinate update sequence $(u_0, u_1, u_2, u_3, \dots )$, we require that each coordinate appears infinitely often. We can assume that $u_0 = i$, since at iteration 0 all the mass is at vertex $i$, thus it is the only vertex to update. Let $S_1$ denote the earliest time after which all of the neighbors of vertex $i$ have been updated at least once:
\[S_1 = \min\{t \geq 1: N_i(1) \subset \{u_0, u_1, \dots u_{t-1}\}\}.\]
This guarantees that $\hat{x}_i^{(S_1)}$ includes the weights from all the length one walks from vertex $i$. We proceed to let $S_2$ denote the earliest time which all vertices within a 2-neighborhood of vertex $i$ have updated once after time $S_1$:
\[S_2 = \min\{t \geq S_1: N_i(2) \subset \{u_{S_1}, u_{S_1 + 1}, \dots u_{t-1}\}\}.\]
This now guarantees that $\hat{x}_i^{(S_2)}$ includes the weights from all the length one and two walks from vertex $i$. We can iteratively define $S_r$ as the time after which the estimate vector has counted and included all walks of length up to $r$:
\[S_l = \min\{t \geq S_{l-1}: N_i(l) \subset \{u_{S_{l-1}}, u_{S_{l-1} + 1}, \dots u_{t-1}\}\}.\]
Since each coordinate appears infinitely often in the sequence, $S_l$ is well defined and finite for all $l$.

Finally we upper bound the error by using a loose upper bound on the weights of all walks with length larger than $l$. By the invariant, it follows that $x_i = \hat{x}_i^{(S_l)} + r^{(S_l)T} x$, which is the sum of all counted or included walks in $\hat{x}_i^{(S_l)}$, and the remaining weight of uncounted walks in $r^{(S_l)T} x$. The weighted sum of all walks of length at most $l$ from vertex $i$ is expressed by $z^T \textstyle\sum_{k=0}^l (G^T)^k e_i$. Thus the error, or weight of uncounted walks, must be bounded by the corresponding weighted sum of the absolute values of the walks of length larger than $l$:
\[-z^T \textstyle\sum_{k=l+1}^{\infty} (\tilde{G}^T)^k e_i \leq r^{(S_l)T} x \leq z^T \textstyle\sum_{k=l+1}^{\infty} (\tilde{G}^T)^k e_i.\]
It follows that 
\[\left|x_i - \hat{x}_i^{(S_l)}\right| \leq z^T \textstyle\sum_{k=0}^{\infty} (\tilde{G}^T)^k (\tilde{G}^T)^{l+1} e_i = x^T (\tilde{G}^T)^{l+1} e_i ,\]
which converges to zero as long as $\rho(\tilde{G}) < 1$.
 \end{proof}

\section{Proof of Theorem \ref{thm:asynch}} \label{sec:pf2}

We provide an analysis for `uniform censored sampling', in which coordinates with nonzero valued current residuals are chosen with equal probability, according to
\begin{align}
\mathcal{P}(u) = \tfrac{\mathbb{I}\left(r_u^{(t)} \neq 0\right)}{\|r^{(t)}\|_0},
\end{align}
where $r^{(t)}$ denotes the current residual after $t$ updates. 

\begin{proof}[Proof of Theorem \ref{thm:asynch}]
Since the algorithm terminates when $\|r^{(t)}\|_2 \leq \epsilon$, it follows from Lemma \ref{lemma:asynch_invariant} that $|\hat{x}_i^{(t)} - x_i| = |r^{(t)T} x| \leq \epsilon \|x\|_2$. Recall that the algorithm chooses a coordinate in each iteration according to the distribution $\mathcal{P}$, as specified in (\ref{eq:sampleDistPhat}). To simplify the analysis, we introduce another probability distribution $\tilde{\mathcal{P}}$, which has a fixed size support of $\min(td, n)$ rather than $\|r^{(t)}\|_0$. We first analyze the convergence of a modified algorithm which samples coordinates according to $\tilde{\mathcal{P}}$. Then we translate the results back to the original algorithm.

Observe that for any $t \in \IntegersP$, there exists a  function $C_t: \Reals^n \rightarrow \{0,1\}^n$, which satisfies the properties that for any $v \in \Reals^n$ and $u = C_t(v)$, if $v_i \neq 0$, then $u_i = 1$, and if $\|v\|_0 \leq t d$, then $\|u\|_0 = \min(t d,n)$. In words, $C_t(v)$ is a function which takes a vector of sparsity at most $td$, and maps it to a binary valued vector which preserves the sparsity pattern of $v$, yet adds extra entries of 1 in order that the sparsity of the output is exactly $\min(td,n)$. We define the distribution $\tilde{\mathcal{P}}$ to choose uniformly at random among the nonzero coordinates of $C_t\left(r^{(t)}\right)$, according to:
\begin{align}
\tilde{\mathcal{P}}(u) = \tfrac{e_u^T S_t\left(r^{(t)}\right)}{\min(td, n)},
\end{align}
where we have suppressed the dependence of $\tilde{\mathcal{P}}$ on $t$ and $r^{(t)}$ for simpler notation. This is a valid probability distribution since for all $t$, $\|r^{(t)}\|_0 \leq td$. We first analyze the asynchronous algorithm which samples coordinates accoridng to $\tilde{\mathcal{P}}$. Lemma \ref{LEMMA:EXPECTEDRESIDUAL} shows that in expectation, the error contracts in each iteration by $(1 - (1-\|G\|_2)/\min(td,n))$. Lemma \ref{LEMMA:SECONDMOMENTRESIDUAL} provides an upper bound on the expected L2-norm of the residual vector $r^{(t)}$. Then we apply Markov's inequality to prove that the algorithm terminates with high probability within a certain number of multiplications.

In order to extend the proofs from $\tilde{\mathcal{P}}$ to $\mathcal{P}$, we define a coupling between two implementations of the algorithm, one which sample coordinates according to  $\tilde{\mathcal{P}}$, and the other which samples coordinates according to $\mathcal{P}$. We prove that in this joint probability space, the implementation which uses distribution $\mathcal{P}$ always terminates in number of iterations less than or equal to the corresponding termination time of the implementation using $\tilde{\mathcal{P}}$. Therefore, computing an upper bound on the number of multiplications required under $\tilde{\mathcal{P}}$ is also an upper bound for the algorithm which uses $\mathcal{P}$.

\begin{lemma}\label{LEMMA:SECONDMOMENTRESIDUAL}
If $\|G\|_2 < 1$, $d \geq 4$, and $n \geq 8$,
\[\E_{\tilde{\mathcal{P}}}\left[\left\|r^{t}\right\|_2^2\right] \leq \min\left( 2 t^{-2(1-\|G\|_2)/d}, 4 e^{-2(t-1) (1 - \|G\|_2)/n}\right).\]
\end{lemma}

By Markov's inequality, $\Prob(\|r^{(t)}\|_2 \geq \epsilon) \leq \delta$ for $\E_{\tilde{\mathcal{P}}}[\|r^{(t)}\|_2^2] \leq \delta \epsilon^2$. Therefore, we can directly apply Lemma \ref{LEMMA:SECONDMOMENTRESIDUAL} to show that if $\|G\|_2 < 1$, $d \geq 4$, and $n \geq 8$, the algorithm terminates with probability at least $1 - \delta$ for 
\begin{align*}
t \geq \min\left(\left(\tfrac{2}{\delta \epsilon^2}\right)^{d/2(1-\|G\|_2)}, 1 + \tfrac{n}{2(1-\|G\|_2)} \ln \left(\tfrac{4}{\delta \epsilon^2}\right) \right).
\end{align*}
Since we are concerned with asymptotic performance, the conditions $d \geq 4$ and $n \geq 8$ are insignificant. To bound the total number of multiplications, we multiply the number of iterations by the maximum degree $d$.

Finally, we complete the proof by translating the analysis for $\tilde{\mathcal{P}}$ to $\mathcal{P}$. Let us consider implementation A, which samples coordinates from $\tilde{\mathcal{P}}$, and implementation B, which samples coordinates from $\mathcal{P}$. Let $R_A$ denote the sequence of residual vectors $r^{(t)}$ derived from implementation A, and let $R_B$ denote the sequence of residual vectors $r^{(t)}$ derived from implementation B. The length of the sequence is the number of iterations until the algorithm terminates. We define a joint distribution such that $\Prob(R_A, R_B) = \Prob(R_A) \Prob(R_B | R_A)$.

Let $\Prob(R_A)$ be described by the algorithm sampling coordinates from $\tilde{\mathcal{P}}$. The sequence $R_A$ can be sampled by separately considering the transitions when non-zero valued coordinates are chosen, and the length of the repeat in between each of these transitions. Given the current iteration $t$ and the sparsity of vector $r^{(t)}$, we can specify the distribution for the number of iterations until the next transition. If we denote $\tau_t = \min\{s : s > t \text{ and } r^{(s)} \neq r^{(t)}\}$, then
\begin{align}
\Prob(\tau_t > k | r^{(t)}) = \textstyle\prod_{q=1}^k \left(1 - \tfrac{\|r^{(t)}\|_0}{\min((t + q) d,n)}\right).
\end{align}
Conditioned on the event that a non-zero valued coordinate is chosen at a particular iteration $t$, the distribution over the chosen coordinate is the same as $\mathcal{P}$.

For all $t$, $r^{(t+1)} \neq r^{(t)}$ if and only if the algorithm chooses a non-zero valued coordinate of $r^{(t)}$ at iteration $t$, which according to $\tilde{\mathcal{P}}$, occurs with probability $1 - \|r^{(t)}\|_0/\min(t d,n)$. Therefore, given the sequence $R_A$, we can identify in which iterations coordinates with non-zero values were chosen. Let $\Prob(R_B | R_A)$ be the indicator function which is one only if $R_B$ is the subsequence of $R_A$ corresponding to the iterations in which a non-zero valued coordinate was chosen.

We can verify that this joint distribution is constructed such that the marginals correctly correspond to the probability of the sequence of residual vectors derived from the respective implementations. For every $(R_A, R_B)$ such that $\Prob(R_B | R_A) = 1$, it also follows that $|R_A| \geq |R_B|$, since $R_B$ is a subsequence. For every $q$,
\begin{align}
&\{(R_A, R_B): |R_A| \leq q\} \subset \{(R_A, R_B): |R_B| \leq q\} \\
&\implies \Prob(|R_A| \leq q) \leq \Prob(|R_B| \leq q).
\end{align}
Therefore, we can conclude that since the probability of the set of realizations such that implementation $A$ terminates within the specified bound is larger than $1 - \delta$, it also follows that implementation $B$ terminates within the specified bound with probability larger than $1 - \delta$. Therefore, since we have proved Theorem \ref{thm:asynch} for implementation $A$, the result also extends to implementation B, i.e., our original algorithm.
 \end{proof}

\section{Proof of Lemma \ref{LEMMA:SECONDMOMENTRESIDUAL}} \label{sec:pf3}

First we prove Lemma \ref{LEMMA:EXPECTEDRESIDUAL}, which shows that a single update task is equivalent in expectation to multiplying the residual vector by the matrix $\left(I - \left(\tfrac{I - G^T}{\min(t d, n)}\right) \right)$.

\begin{lemma} \label{LEMMA:EXPECTEDRESIDUAL}
If $\|G\|_2 < 1$, for all $t$,
\begin{flalign*}
~~~~&\text{\normalfont (a) } \E_{\tilde{\mathcal{P}}}\left[\left.r^{(t+1)} \right| r^{(t)}\right] = \left(I - \left(\tfrac{I - G^T}{\min(t d, n)}\right) \right) r^{(t)}, &\\
~~~~&\text{\normalfont (b) } \left\|\E_{\tilde{\mathcal{P}}}\left[r^{(t)}\right]\right\|_2 \leq \min\left(t^{- (1 - \|G\|_2)/d}, e^{- (t-1)(1-\|G\|_2)/n}\right). &
\end{flalign*}
\end{lemma}

\begin{proof}[Proof of Lemma \ref{LEMMA:EXPECTEDRESIDUAL}]
We will use induction to get an expression for $\E_{\tilde{\mathcal{P}}}\left[r^{(t)}\right]$. Recall that $r^{(0)} = e_i$. Since there is only a single coordinate to choose from, $r^{(1)}$ is also predetermined, and is given by $r^{(1)} = G^T e_i$.
\begin{align}
\E_{\tilde{\mathcal{P}}}\left[\left.r^{(t+1)} \right| r^{(t)}\right] &= r^{(t)} - (I - G^T) \textstyle\sum_u \tilde{\mathcal{P}}(u) e_u r_u^{(t)}. \label{eq:intermedCondExp}
\end{align}
By design of $\tilde{\mathcal{P}}$, we know that $\tilde{\mathcal{P}}(u) = 1/\min(td,n)$ for all $u$ such that $r_u^{(t)} \neq 0$. Therefore,
\begin{align}
\textstyle\sum_u \tilde{\mathcal{P}}(u) e_u r_u^{(t)} = \tfrac{r_u^{(t)}}{\min(td,n)}.
\end{align}
We substitute this into (\ref{eq:intermedCondExp}) to show that
\begin{align}
\E_{\tilde{\mathcal{P}}}\left[\left.r^{(t+1)} \right| r^{(t)}\right] &= \left(I - \left(\tfrac{I - G^T}{\min(t d, n)}\right) \right) r^{(t)}.
\end{align}
Using the initial conditons $r^{(1)} = G^T e_i$ and the law of iterated expectation, it follows that
\begin{align}
\E_{\tilde{\mathcal{P}}}\left[r^{(t)T}\right] &= e_i^T G \textstyle\prod_{k=1}^{t-1} \left(I - \left(\tfrac{I - G}{\min(k d, n)}\right) \right).
\end{align}
Therefore,
\begin{align}
&\left\|\E_{\tilde{\mathcal{P}}}\left[r^{(t)}\right]\right\|_2 \nonumber \\
&\leq \|G\|_2 \textstyle\prod_{k=1}^{t-1} \left(1 - \left(\tfrac{1 - \|G\|_2}{\min(k d, n)}\right) \right), \nonumber \\
&\leq \|G\|_2 \exp\left( - \textstyle\sum_{k=1}^{t-1} \tfrac{1 - \|G\|_2}{\min(k d,n)} \right), \nonumber \\
&\leq \|G\|_2 \min\left(\exp\left( - \textstyle\sum_{k=1}^{t-1} \tfrac{1 - \|G\|_2}{k d} \right), \exp\left( - \textstyle\sum_{k=1}^{t-1} \tfrac{1 - \|G\|_2}{n} \right)\right).
\end{align}
Since $\|G\|_2 < 1$ by assumption, and using the property that $\textstyle\sum_{k=1}^{t-1} \tfrac{1}{k} > \ln(t)$, it follows that
\begin{align*}
\left\|\E_{\tilde{\mathcal{P}}}\left[r^{(t)}\right]\right\|_2 &\leq \min\left(t^{- (1 - \|G\|_2)/d}, e^{- (t-1)(1-\|G\|_2)/n}\right).
\end{align*}
 \end{proof}

We use Lemma \ref{LEMMA:EXPECTEDRESIDUAL} to prove Lemma \ref{LEMMA:SECONDMOMENTRESIDUAL}.

\begin{proof}[Proof of Lemma \ref{LEMMA:SECONDMOMENTRESIDUAL}]
Observe that
\begin{align}
&\E_{\tilde{\mathcal{P}}}\left[\left\|r^{(t+1)}\right\|_2^2\right] \nonumber \\
&= \E_{\tilde{\mathcal{P}}}\left[\left\|r^{(t+1)} - \E_{\tilde{\mathcal{P}}}\left[r^{(t+1)}\right]\right\|_2^2 \right] + \left\|\E_{\tilde{\mathcal{P}}}\left[r^{(t+1)}\right]\right\|_2^2, \label{eq:resSecMomnent}
\end{align}
and 
\begin{align}
&\E_{\tilde{\mathcal{P}}}\left[\left.\left\|r^{(t+1)} - \E_{\tilde{\mathcal{P}}}\left[r^{(t+1)}\right]\right\|_2^2 \right| r^{(t)}\right] \nonumber \\
&= \E_{\tilde{\mathcal{P}}}\left[\left. \left\|r^{(t+1)}\right\|_2^2 \right| r^{(t)}\right] - \left\|\E_{\tilde{\mathcal{P}}}\left[\left.r^{(t+1)}\right| r^{(t)}\right] \right\|_2^2. \label{eq:cond_resDiff}
\end{align}
Based on the update equation 
\[r^{(t+1)} = r^{(t)} - (I - G^T) e_u r_u^{(t)},\]
we can compute that
\begin{align}
&\E_{\tilde{\mathcal{P}}}\left[\left.\left\|r^{(t+1)}\right\|_2^2 \right| r^{(t)}\right] \nonumber \\ 
&= \textstyle\sum_u \tilde{\mathcal{P}}(u) (r^{(t)} - (I - G^T) e_u r_u^{(t)})^T (r^{(t)} - (I - G^T) e_u r_u^{(t)}), \nonumber \\
&= r^{(t)T} r^{(t)} - \left(\textstyle\sum_u \tilde{\mathcal{P}}(u) r_u^{(t)} e_u^T\right) (I - G) r^{(t)} \nonumber \\
&~~~~ - r^{(t)T} (I - G^T) \left(\textstyle\sum_u \tilde{\mathcal{P}}(u) e_u r_u^{(t)}\right) \nonumber \\
&~~~~ + \textstyle\sum_u \tilde{\mathcal{P}}(u) r_u^{(t) 2} \left[(I - G) (I - G^T)\right]_{uu}. \label{eq:condSecMomComp}
\end{align}
By the design, $\tilde{\mathcal{P}}(u) r_u^{(t)} = r_u^{(t)} / \min(td,n)$, so that
\begin{align}
\textstyle\sum_u \tilde{\mathcal{P}}(u) e_u r_u^{(t)} = \tfrac{r_u^{(t)}}{\min(td,n)}. \label{eq:prob1}
\end{align}
Similarly, since $\tilde{\mathcal{P}}(u) r_u^{(t)2} = r_u^{(t)2} / \min(td,n)$,
\begin{align}
\textstyle\sum_u \tilde{\mathcal{P}}(u) r_u^{(t) 2} \left[(I - G) (I - G^T)\right]_{uu}
= \tfrac{r^{(t)T} D r^{(t)}}{\min(td,n)}, \label{eq:prob2}
\end{align}
where $D$ is defined to be a diagonal matrix such that
\begin{align}
D_{uu} &= \left[(I - G) (I - G^T)\right]_{uu} 
= 1 - 2G_{uu} + \textstyle\sum_k G_{uk}^2.
\end{align}
Therefore, we substitute (\ref{eq:prob1}) and (\ref{eq:prob2}) into (\ref{eq:condSecMomComp}) to show that
\begin{align}
\E_{\tilde{\mathcal{P}}}&\left[\left.\left\|r^{(t+1)}\right\|_2^2 \right| r^{(t)}\right] 
= r^{(t)T} \left(I - \tfrac{2I - G - G^T - D}{\min(t d, n)}\right) r^{(t)}. \label{eq:cond_resSecondMom}
\end{align}
We substitute (\ref{eq:cond_resSecondMom}) and Lemma \ref{LEMMA:EXPECTEDRESIDUAL}a into (\ref{eq:cond_resDiff}) to show that
\begin{align}
\E_{\tilde{\mathcal{P}}}&\left[\left.\left\|r^{(t+1)} - \E_{\tilde{\mathcal{P}}}\left[r^{(t+1)}\right]\right\|_2^2 \right| r^{(t)}\right], \nonumber \\
&= r^{(t)T} \left(I - \tfrac{2I - G - G^T - D}{\min(t d, n)}\right) r^{(t)} \nonumber \\
&~~~~ - r^{(t)^T}\left(I - \left(\tfrac{I - G}{\min(t d, n)}\right) \right) \left(I - \left(\tfrac{I - G^T}{\min(t d, n)}\right) \right) r^{(t)}, \nonumber \\
&= r^{(t)T} \left( \tfrac{D}{\min(t d, n)} - \tfrac{(I - G) (I - G^T)}{\min(t d, n)^2} \right) r^{(t)}, \nonumber \\
& \leq\left\| \tfrac{D}{\min(t d, n)} - \tfrac{(I - G) (I - G^T)}{\min(t d, n)^2} \right\|_2 \left\|r^{(t)}\right\|_2^2. \label{eq:cond_resDiff_expr}
\end{align}
By definition, for all $u$,
\begin{align}
\|G\|_2 &= \left\|G^T\right\|_2 = \max_{\|x\|_2 = 1} \left\|G^T x\right\|_2 \\
&\geq \sqrt{e_u^T G G^T e_u} = \sqrt{\textstyle\sum_k G_{uk}^2}.
\end{align}
Therefore, $G_{uu}^2 \leq \textstyle\sum_k G_{uk}^2 \leq \left\|G\right\|_2^2$, and 
\begin{align}
D_{uu} &= 1 - 2G_{uu} + \textstyle\sum_k G_{uk}^2 \\
&\leq 1 + 2 \|G\|_2 + \|G\|_2^2 \\
&= (1 + \|G\|_2)^2.\label{eq:D_bound}
\end{align}
Substitute (\ref{eq:D_bound}) into (\ref{eq:cond_resDiff_expr}) to show that
\begin{align}
\E_{\tilde{\mathcal{P}}}&\left[\left.\left\|r^{(t+1)} - \E_{\tilde{\mathcal{P}}}\left[r^{(t+1)}\right]\right\|_2^2 \right| r^{(t)}\right]  \nonumber \\
&\leq \tfrac{(1 + \|G\|_2)^2}{\min(t d, n)} \left(1 + \tfrac{1}{\min(t d, n)}\right) \left\|r^{(t)}\right\|_2^2. \label{eq:cond_resDiff_expr_clean}
\end{align}
We will use the two expressions given in Lemma \ref{LEMMA:EXPECTEDRESIDUAL}b to get different upper bounds on $\E_{\tilde{\mathcal{P}}}[\|r^{(t+1)}\|_2^2]$, and then take the minimum. The first bound is most relevant in the sparse setting when $n$ is large and $d$ and $\|G\|_2$ are small. We substitute (\ref{eq:cond_resDiff_expr_clean}) and the first expression in Lemma \ref{LEMMA:EXPECTEDRESIDUAL}b into (\ref{eq:resSecMomnent}) to show that 
\begin{align}
\E_{\tilde{\mathcal{P}}}[\|r^{(t+1)}\|_2^2] \leq a_t \E_{\tilde{\mathcal{P}}}[\|r^{(t)}\|_2^2] + b_t,
\end{align}
for
\begin{align}
a_t &= \tfrac{(1 + \|G\|_2)^2}{\min(t d, n)} \left(1 + \tfrac{1}{\min(t d, n)}\right),
\end{align}
and 
\begin{align}
b_t &= (t+1)^{- 2(1 - \|G\|_2)/d}.
\end{align}
Therefore, $\E_{\tilde{\mathcal{P}}}[\|r^{(t+1)}\|_2^2] \leq \textstyle\sum_{k=1}^t Q_k$ for
\begin{align}
Q_k &= \left(\textstyle\prod_{m=k+1}^t a_m \right) b_k \\
&= \left(\textstyle\prod_{m=k+1}^t \tfrac{(1 + \|G\|_2)^2}{\min(m d, n)} \left(1 + \tfrac{1}{\min(m d, n)}\right) \right) \nonumber \\
&~~~~~ \cdot (k+1)^{- 2 (1 - \|G\|_2)/d}.
\end{align}
The ratio between subsequent terms can be upper bounded by
\begin{align}
&\tfrac{Q_{k}}{Q_{k+1}} \leq \tfrac{(1 + \|G\|_2)^2}{\min((k+1) d, n)} \left(1 + \tfrac{1}{\min((k+1) d, n)}\right)   \left(\tfrac{k+1}{k+2}\right)^{- 2 (1 - \|G\|_2)/d}.
\end{align}
For $k \geq 1$, $d \geq 4$, and $n \geq 8$, 
\begin{align}
\tfrac{Q_k}{Q_{k+1}} \leq \tfrac{4}{8}\left(1 + \tfrac{1}{8}\right) \left(\tfrac{2}{3}\right)^{2/4} < \tfrac{1}{2}.
\end{align}
It follows that
\begin{align}
\E_{\tilde{\mathcal{P}}}\left[\left\|r^{(t+1)}\right\|_2^2\right] &\leq Q_t \textstyle\sum_{k=1}^t \left(\tfrac{1}{2}\right)^{t-k} \\
&\leq 2 (t+1)^{-2(1-\|G\|_2)/d}.
\end{align}

We similarly obtain another bound by using the second expression of Lemma \ref{LEMMA:EXPECTEDRESIDUAL}b. This bound applies in settings when the residual vector $r^{(t)}$ is no longer sparse. By Lemma \ref{LEMMA:EXPECTEDRESIDUAL}b,
\[\E_{\tilde{\mathcal{P}}}[\|r^{(t+1)}\|_2^2] \leq a_t \E_{\tilde{\mathcal{P}}}[\|r^{(t)}\|_2^2] + b'_t,\]
for $b'_t = e^{- 2t(1-\|G\|_2)/n}$.
Therefore, $\E_{\tilde{\mathcal{P}}}[\|r^{(t+1)}\|_2^2] \leq \textstyle\sum_{k=1}^t Q'_k$ for
\begin{align}
Q'_k = \left(\textstyle\prod_{m=k+1}^t \tfrac{(1 + \|G\|_2)^2}{\min(m d, n)} \left(1 + \tfrac{1}{\min(m d, n)}\right) \right) e^{-2k(1 - \|G\|_2)/n}. \nonumber 
\end{align}
The ratio between subsequent terms can be upper bounded by
\begin{align}
\tfrac{Q'_{k}}{Q'_{k+1}} \leq \tfrac{(1 + \|G\|_2)^2 e^{2(1-\|G\|_2))/n}}{\min((k+1) d, n)} \left(1 + \tfrac{1}{\min((k+1) d, n)}\right). \nonumber
\end{align}
For $k \geq 1$, $d \geq 4$, and $n \geq 8$, 
\begin{align}
\tfrac{Q'_k}{Q'_{k+1}} \leq \tfrac{9 e^{2(1-\|G\|_2)/n}}{16} < \tfrac{3}{4}.
\end{align}
It follows that
\begin{align}
\E_{\tilde{\mathcal{P}}}[\|r^{(t+1)}\|_2^2] \leq Q'_t \textstyle\sum_{k=1}^t \left(\tfrac{3}{4}\right)^{t-k}
\leq 4 e^{-2t(1 - \|G\|_2)/n}.
\end{align}
 \end{proof}



%


\ifCLASSOPTIONcompsoc
  \section*{Acknowledgments}
\else
  \section*{Acknowledgment}
\fi

This work is supported in parts by ARO under MURI award W911NF-11-1-00365, by AFOSR under MURI award FA9550-09-1-0538, by ONR under the Basic Research Challenge No. N000141210997,  by DARPA under grant W911NF-16-1-055, and by NSF under grants CNS-1161964, CMMI-1462158, CMMI-1634259 and a Graduate Fellowship.

\ifCLASSOPTIONcaptionsoff
  \newpage
\fi



%

\bibliographystyle{IEEEtran}
\bibliography{AxbBibliography}

\vfill



\end{document}